\documentclass[]{JHEP3}

\usepackage{graphicx}
\usepackage{epsfig}
\usepackage{amssymb}
\usepackage{subfigure}

\newcommand{\myref}[1]{(\ref{#1})}
\newcommand{\smartpap}{p\hskip-7pt\hbox{$^{^{(\!-\!)}}$}}

\title{
Precision electroweak calculation of the charged current Drell-Yan process
}
\author{C.M.~Carloni Calame \\ 
INFN, Sezione di Pavia and Dipartimento di Fisica Nucleare e Teorica, 
Universit\`a di Pavia,
Via A.Bassi 6, I-27100 Pavia, Italy\\
Email: \email{Carlo.Carloni.Calame@pv.infn.it} 
}
\author{G.~Montagna \\
Dipartimento di Fisica Nucleare e Teorica, 
Universit\`a di Pavia and
INFN, Sezione di Pavia, Via A.Bassi 6, I-27100 Pavia, Italy \\
Email: \email{Guido.Montagna@pv.infn.it}
}
\author{ O.~Nicrosini \\
INFN, Sezione di Pavia, Via A.Bassi 6, I-27100 Pavia, Italy \\
Email: \email{ Oreste.Nicrosini@pv.infn.it}
}
\author{ A.~Vicini \\
Dipartimento di Fisica, Universit\`a degli Studi di Milano and
INFN, Sezione di Milano,
Via Celoria 16, I--20133 Milano, Italy \\ 
Email: \email{Alessandro.Vicini@mi.infn.it}
}


\abstract{
We present a detailed study of the charged current Drell-Yan process, which
includes the exact  ${\cal O}(\alpha)$ electroweak corrections
properly matched with leading-log effects
due to multiple-photon emission, as required by the experiments at the Tevatron
and the LHC. Numerical results for the relevant observables of single $W$ boson
production at hadron colliders are presented. The impact of the
radiative corrections
and of some sources of theoretical uncertainty is discussed in detail.
The calculation
has been implemented in the new version of the event generator
$\tt HORACE$, which is available for precision simulations of the
charged current Drell-Yan process.
}

        \preprint{
         FNT/T 2006/08\\
         IFUM-874/FT }

\begin{document}

\newcommand{\be}{\begin{equation}}
\newcommand{\ee}{\end{equation}}
\newcommand{\nn}{\nonumber}
\newcommand{\bea}{\begin{eqnarray}}
\newcommand{\eea}{\end{eqnarray}}
\newcommand{\bfig}{\begin{figure}}
\newcommand{\efig}{\end{figure}}
\newcommand{\bc}{\begin{center}}
\newcommand{\ec}{\end{center}}
\def\ad{\dot{\alpha}}
\def\ov{\overline}
\def\hlf{\frac{1}{2}}
\def\qrt{\frac{1}{4}}
\def\as{\alpha_s}
\def\at{\alpha_t}
\def\ab{\alpha_b}
\def\sq2{\sqrt{2}}
\newcommand{\smallz}{{\scriptscriptstyle Z}} %
\newcommand{\mz}{m_\smallz}
\newcommand{\smallw}{{\scriptscriptstyle W}}
\newcommand{\mw}{m_\smallw} 
\newcommand{\gw}{\Gamma_{\smallw}} 
\newcommand{\sw}{s_{\smallw}} 
\newcommand{\cw}{c_{\smallw}} 
\newcommand{\sdw}{\sin^2\theta_{\smallw}} 
\newcommand{\cdw}{\cos^2\theta_{\smallw}} 
\newcommand{\sqw}{\sin^4\theta_{\smallw}} 
\newcommand{\smallh}{{\scriptscriptstyle H}}
\newcommand{\mh}{m_\smallh}
\newcommand{\mt}{m_t}
\newcommand{\wh}{w_\smallh}
\newcommand{\oa}{${\cal O}(\alpha)~$} 
\newcommand{\oab}{${\cal O}(\alpha)$} 
\def\th{t_\smallh}
\def\zh{z_\smallh}
\newcommand{\Mvariable}[1]{#1}
\newcommand{\Mfunction}[1]{#1}
\newcommand{\Muserfunction}[1]{#1}
%
%


\renewcommand{\thefootnote}{\fnsymbol{footnote}}

%
%
\setcounter{footnote}{0}
\section{Introduction}
At hadron colliders, such as the Fermilab Tevatron and the CERN LHC,
the production of a high transverse momentum lepton pair,
known as Drell-Yan process \cite{DY}, plays an important role:
it allows, in the charged current channel,
a high precision determination of two fundamental parameters
of the Standard Model, namely the mass and the decay width of the $W$
boson \cite{CDF}; it provides, both in neutral and charged current channels,
stringent constraints on the density functions which
describe the partonic content of the proton \cite{pdf};
it can be used as a standard reference process 
and therefore as a luminosity monitor of the collider
\cite{DPZ,FM}. Furthermore, it
represents a background to the search for new heavy gauge bosons
\cite{CDFnewgauge}.

The accuracy in the determination of the theoretical cross section has
greatly increased over the years. The calculation of next-to-leading
order (NLO) QCD
corrections~\cite{AEM}
has been one of the first test grounds of perturbative QCD.
Next-to-next-to-leading order (NNLO)
QCD corrections to the total cross section have been computed
in ref. \cite{HvNM}, but  differential distributions with the same accuracy
have been obtained only recently in ref.\cite{ADMP}.
The size of the NNLO QCD corrections and the improved stability of the
results against changes of the renormalization/factorization scales
raises the question of the relevance of the \oa electroweak (EW) radiative
corrections, which were computed, in the charged current channel,
first in the pole approximation~\cite{HW,BKW} and then fully in
refs. \cite{ZYK,DK,BW,SANC}.

A realistic phenomenological study and the data analysis require
the inclusion of the relevant radiative corrections 
and their implementation into Monte Carlo event generators,
in order to simulate all the experimental cuts 
and to allow, for instance, an accurate determination of the 
detector acceptances.
The Drell-Yan processes are included in the standard QCD Parton Shower
generators $\tt HERWIG$ and $\tt PYTHIA$~\cite{HERWIG,PYTHIA}.
Recently there have been important progresses to improve the QCD
radiation description to NLO, which has been implemented in 
the code $\tt MC@NLO$~\cite{MC@NLO}.
Another important issue is the good description of the 
intrinsic transverse momentum of the gauge boson, which can be
obtained by resumming up to all orders the contributions of the form
$\alpha_s\log(p_\perp^{\scriptscriptstyle W}/\mw)$.
The generator $\tt RESBOS$~\cite{BY}, used for data analysis at
Tevatron, includes these effects.

If the inclusion of QCD radiation is mandatory for the simulation of
any process at a hadron collider, one should not neglect the impact of
EW corrections on the precision measurement of some
Standard Model (SM) observables, 
like the $W$ boson mass and decay width. For instance,
the generator
$\tt WGRAD$~\cite{BW} includes the exact \oa EW corrections, which have been
shown to induce a shift on the value of $\mw$ extracted from the
Tevatron data of about 160 MeV in the muon channel~\cite{CDF},
mostly due to final-state QED radiation.
In view of the very high experimental precision foreseen at the LHC
($\Delta\mw \approx 15$~MeV),
final-state higher-order (beyond \oab) QED corrections may induce a
significant shift, as shown in ref.~\cite{CMNT}. 
Some event generators
can account also for multiple-photon radiation: in the published
version of $\tt HORACE$~\cite{CMNT,programma} final-state QED radiation was
simulated by means of a QED Parton Shower \cite{CarloPS}; the generator 
$\tt WINHAC$~\cite{WINHAC} uses the Yennie-Frautchi-Suura \cite{YFS}
formalism to
exponentiate final-state-like EW \oa corrections; finally, the
standard tool $\tt PHOTOS$~\cite{PHOTOS} can be used to describe QED
radiation in the $W$ decay.

A first attempt to study the combined effect of EW and QCD corrections
has been presented in ref. \cite{CY}.

The predictions of $\tt WINHAC$ and $\tt HORACE$ and of
$\tt WINHAC$ and $\tt PHOTOS$, for $W$-decay,
have been compared in the papers of ref.~\cite{PScomparison}.
A detailed series of tuned comparisons between different EW Monte
Carlo generators have been done in ref.~\cite{LesHouches},
in order to check the reliability of different numerical predictions,
with \oa accuracy,
in a given setup of input parameters and cuts.

Since the Drell-Yan events can be used, in principle, to determine the collider
luminosity at a few per cent level, the theoretical cross section
must be known with the same accuracy, requiring also the inclusion
of \oa EW corrections. Furthermore, the \oa EW contributions give
large corrections to the tails of the transverse mass and lepton
tranverse momentum distributions, because of the presence of large EW
Sudakov logarithms \cite{DK,BW}. These regions are important for the  
search of new heavy gauge bosons.


The aim of this paper is to present a precision EW calculation of the
charged current Drell-Yan process, 
which includes the exact \oa EW matrix elements properly matched with
leading-logarithmic  higher-order QED corrections in the
Parton Shower approach.
The matching here presented of perturbative corrections with 
Parton Shower, 
which is a topic of great interest in modern QCD simulations
\cite{NS},
is the first example of such an application in the EW sector and is
realized along 
the lines already presented in ref. \cite{BCMNP}. Several distributions of
physical interest are analyzed,
disentangling the effect of different classes 
of radiative corrections and discussing various sources of theoretical
uncertainty.
The calculation is implemented in the Monte Carlo event generator $\tt
HORACE$,
which combines, in a unique tool,
the good features of the QED Parton Shower approach with the 
additional effects
present in the exact \oa EW calculation.
This task is non trivial from several technical points of view and
faces all the conceptual problems of developing a NLO event generator.

The paper is organised as follows.
In Section~\ref{oa} we present the calculation 
of the \oa EW corrections to the partonic process
$u \bar d \to l^+ \nu_l (\gamma)$.
In Section~\ref{matching} we describe the matching of the fixed order
results with the QED Parton Shower.
In Section~\ref{hadronic} we present the computation of the hadron-level
cross section $\sigma\left(p\smartpap \to l^+\nu_l (n\gamma)\right)$
and discuss the subtraction of the initial-state collinear
singularities to all orders.
In Section~\ref{results}
we present phenomenological results for several physical distributions
and discuss the impact of EW \oa and of higher-order QED
corrections. Finally, in Section~\ref{concllabel} we draw our
conclusion and discuss possible developments of this work.

\section{Partonic process: matrix elements calculation}
\label{oa}
\subsection{Born approximation}
We consider the charged current Drell-Yan partonic process
$u(p_1)~ {\bar d(p_2)} \to \nu_l(p_3)~ l^+(p_4)$.
This process is a weak charged current process and its amplitude,
in unitary gauge, is proportional, at tree level,
to the square of the $SU(2)_L$ coupling constant
$g$:
\be
{\cal M}_0 = i~\frac{g^2 V_{ud}}{2}
~\frac{g_{\mu\nu} - k_\mu k_\nu/\mw^2}{s-\mw^2+i\gw\mw} 
\left[
{\bar v}(p_2) \gamma^{\mu} \frac{1-\gamma_5}{2} u(p_1)
\right]
\left[
{\bar u}(p_3) \gamma^{\nu} \frac{1-\gamma_5}{2} v(p_4)
\right]
\label{M0}
\ee
where $V_{ud}$ is the CKM matrix-element,
$\mw$ is the $W$-boson mass and $\gw$ is the $W$ decay width,
necessary to describe the $W$ resonance region,
$s=(p_1+p_2)^2$ is the squared center-of-mass energy and
$k^\mu=p_1^\mu+p_2^\mu$.
The differential cross section, mediated over initial-state and
summed over final-state spins, mediated over initial-state colours and
in the limit of vanishing fermion masses,
is given by
\be
\frac{d\sigma_0}{d\Omega} = 
\frac{g^4~|V_{ud}|^2}{768 \pi^2}~
\frac{1}{(s-\mw^2)^2+\gw^2\mw^2}~
\frac{u^2}{s}
\label{Borndiff}
\ee
where $u=(p_1-p_4)^2$.
The weak coupling $g$ can be expressed in terms of
$\alpha$, the fine structure constant,
and of $\sin\theta_{\smallw}$, 
the sinus of the weak mixing angle, 
via the relation $4 \pi\alpha = g^2 \sin^2\theta_{\smallw}$.
The weak mixing angle is defined as 
$\cos\theta_{\smallw}\equiv\mw/\mz$, where $\mz$ is
the $Z$ boson mass.


\subsection{The \oa calculation}
The complete EW \oa corrections to the charged current Drell-Yan process
have already been computed in refs. \cite{HW,BKW,ZYK,DK,BW,SANC}.
We have repeated independently the calculation and we summarize here
its main features.

The \oa corrections include the contribution of real and virtual corrections.
The virtual corrections follow from the perturbative expansion
of the $2\to 2$ scattering amplitude
${\cal M} = {\cal M}_0 + {\cal M}_{\alpha}^{virt} + \cdots$ and contribute,
at \oab, with $2 {\mathrm Re}({\cal M}_{\alpha}^{virt} {\cal M}_0^*)$.  
The \oa virtual amplitude includes two contributions, namely the 
one-loop renormalization of the tree-level amplitude and  the
virtual one-loop diagrams. The real corrections are due to the
emission of one extra real photon and represent the lowest order of
the radiative process
$u(p_1) {\bar d(p_2)} \to \nu_l(p_3) l^+(p_4) \gamma(k)$.
They can be further divided in soft and hard corrections,
${\cal M}_{1}={\cal M}_{1}^{soft}+{\cal M}_{1}^{hard}$.
The former respect, by definition, the Born-like $2\to 2$ kinematics
and can be factorized as 
$|{\cal M}_{1}^{soft}|^2 = \delta_{SB} |{\cal M}_0|^2$
where $\delta_{SB}$ is a universal factor that depends only on the
external particles. The total cross section includes soft and hard corrections
and is independent of the
cut-off used to define the two energy regions.
Virtual and real soft corrections are separately divergent due to the
emission of soft photons, but the divergence cancels in the sum of the
two contributions.

\subsubsection{Virtual corrections}
\begin{figure}
\begin{center}
\includegraphics[scale=0.8]{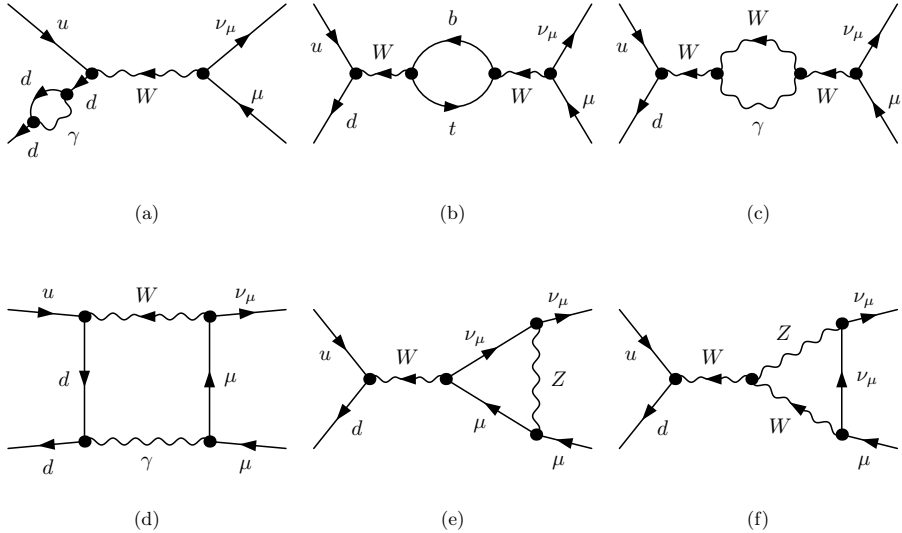}
\caption{Some examples of one-loop virtual diagrams.}
\label{figvirt}
\end{center}
\end{figure}

The \oa virtual corrections to a $2\to 2$ reaction include
contributions of counterterm, self-energy, vertex and box corrections.
Few diagrams representative of the different kinds 
of corrections are depicted in figure~\ref{figvirt}
and have been calculated using the packages $\tt FeynArts$ and
$\tt FormCalc$~\cite{FA,LT}.
The numerical evaluation of the 1-loop integrals has been done using
the package $\tt LoopTools2$~\cite{LT}, based on the library
$\tt ff$~\cite{ff}.
We will write the 1-loop virtual amplitude as
${\cal M}_{\alpha}^{virt}=
{\cal M}_{\alpha}^{cts}+
{\cal M}_{\alpha}^{self}+
{\cal M}_{\alpha}^{vertex}+
{\cal M}_{\alpha}^{box}$.
We have explicitly checked that our results numerically agree with those
of refs.~\cite{DK,SANC}; the fermion mass corrections, which
are present in our calculation and neglected in refs.~\cite{DK,SANC},
turn out to be negligible from a numerical point of view.
The mass of the fermions in the scalar 1-loop
integrals regularizes in a natural way the mass singularities due to the
emission of a (virtual) collinear photon.
The infrared divergence of the integrals has been regularized by means
of a small photon mass $\lambda$.

The introduction of the $W$ decay width in the propagator of the $W$
boson is mandatory to describe the resonance region and to regularize the
divergence due to the pole of the propagator. In order to account for
the $W$ width, we observe that
the $W$ propagator, in the 't Hooft-Feynman gauge, 
can be modified at 1-loop in the following way:
\bea
&&(-i g^{\mu\nu})\frac{1}{s-\mw^2+i\gw\mw}
\to\nonumber\\
&&(-i g^{\mu\nu})\frac{1}{s-\mw^2+i\gw\mw}
\left(\Pi_{WW}(s)+\delta\mw^2+(s-\mw^2)\delta Z_\smallw\right)
\frac{1}{s-\mw^2}
\label{Wprop}
\eea
where $\Pi_{WW}(s)$ is the transverse part of the $W$ self-energy corrections,
$\delta\mw^2$ and $\delta Z_\smallw$ are respectively the $W$ mass
and wave function renormalization constants.
The two counterterms cancel the divergences present in the self-energy
corrections. We remark that the second factor $1/(s-\mw^2)$ in
eq.~\myref{Wprop} is not corrected
by the decay width, to avoid double counting; 
we can check, by expanding the self-energy corrections about
$s=\mw^2$, that the \oa expression is regular for $s\to\mw^2$.
The contemporary presence of the decay width in the propagator and of
the explicit \oa corrections does not yield any double counting~\cite{DK}: in
fact the imaginary part of the self-energy corrections contributes
only at the 2-loop level to the cross section and does not enter in
the 1-loop virtual contribution
$2 {\mathrm Re} \left({\cal M}_\alpha^{virt} {\cal  M}_0^*\right)$.
The $W$ self-energy includes also contributions from diagrams having a
$W$ and a photon running in the loop, which develop a logarithmic
divergence when $s=\mw^2$. In the 1-loop virtual amplitude, the
coefficient of $\log(s-\mw^2)$ is gauge invariant; we can therefore
regularize the logarithmic divergence by replacing
$\log(s-\mw^2) \to \log(s-\mw^2+i\gw\mw)$ in all the self-energy, vertex
and box scalar integrals, without spoiling the gauge-invariance of the
calculation.

Examples of vertex and box corrections are depicted in figure~\ref{figvirt}.
The vertex diagrams with a trilinear gauge boson vertex 
and the box diagrams with a photon in the loop yield the 
logarithmic divergence at $s=\mw^2$ previously discussed.
The abelian vertex diagrams and the box diagrams with a photonic correction
are infra-red divergent.
All the vertex and box diagrams with a $Z$ boson exchange
yield the so-called EW Sudakov logarithms, namely terms like 
$\alpha\log^2\left(s/\mz^2 \right)$, whose importance grows for large
invariant mass of the final state lepton pair, while they are almost
negligible at the $W$ resonance.

The calculation has been repeated with two different gauge choices,
namely the $R_{\xi}$ with $\xi=1$ gauge and the background field
gauge, with parameter $q=1$~\cite{BFG}.
The two results
perfectly agree, and this is an important check on the calculation of
the bosonic self-energy and of the non-abelian vertex corrections.

Concerning the renormalization of the 1-loop amplitude,
the UV divergences which appear from the
virtual diagrams can be cancelled with the mass, $\delta\mw^2$, and wave
function, $\delta Z_{\smallw}$, renormalization constants of the $W$ boson and 
by the renormalization of the two vertices $Wu{\bar d}$ and $Wl^+\nu_l$. 
The latter include the charge renormalization and the wave
function renormalization of the external fermions and of the $W$ boson.
Being the tree-level amplitude of eq.~\myref{M0} proportional to $g_0^2$, 
using the electric charge and the gauge boson masses $(e,\mw,\mz)$ 
as input parameters, we write the bare coupling  $g_0=e_0/s_{\smallw,0}$
and then replace it in terms of renormalized quantities and of
counterterms $g_0 = e/\sw ~(1-\delta e/e)/(1-\delta \sw/\sw)$.
The electric charge counterterm is fixed by the request that in
Thomson scattering the renormalized charge is given by the fine
structure constant;
its expression depends on the quark masses running in the photon
vacuum polarization, the
value of which can be adjusted in order to make the running
electric charge reproduce the value $\alpha(\mz^2)$~\cite{fred}.
The weak mixing angle counterterm is given as a combination of
the mass counterterms of the $W$ and $Z$ bosons, following from its
definition.
The $W$ boson mass and wave function renormalization constants are
defined in the on-shell scheme.

The choice of the input parameters of the SM lagrangian has an impact
on the prediction of the physical observables.
If ideally one were able to resum exactly the perturbative expansion,
the predictions would be the same in any scheme.
On the other hand the truncation of the perturbative expansion
induces a dependence on the scheme, which is formally of higher order,
but which can be numerically relevant.

Some of the possible options in the gauge sector of the EW SM are:
i)~$\alpha, G_{\mu}, \mz$, ii)~$\alpha, \mw, \mz$, 
iii)~$G_{\mu}, \mw, \mz$.
The first option, used for LEP1 analyses, is based on the best measured
EW quantities and minimizes the parametric dependence of the
predictions on the inputs.
In this input scheme, the value of $\mw$ is a predicted quantity.
For $W$ physics at hadron colliders, the other schemes are
preferable because $\mw$ is an input parameter.

The second choice ($\alpha(0)$-scheme) has the proper coupling
$\alpha$ for the real photon emission diagrams and parametrizes the
charged current coupling $g$ as $\sqrt{4\pi\alpha}/\sw$.

The choice which seems to be more natural for $W$ physics is the third one
($G_{\mu}$ scheme), where
the weak coupling $g$ is related to the Fermi constant and to the $W$
boson mass by the relation
\be
\frac{G_{\mu}}{\sqrt{2}}~=~\frac{g^2}{8\mw^2}\left(1+\Delta r\right)
\label{gmu}
\ee
The quantity $\Delta r$ represents all the radiative corrections to the
muon-decay amplitude~\cite{Sirlin80}.
Introducing
\be
g^2=4 \sqrt{2} G_{\mu} \mw^2 (1-\Delta r)
\label{gdr}
\ee
in the tree level amplitude~\myref{M0} we generate an additional
contribution to the \oa correction proportional to $\Delta r$.
Being the vertex between charged particles and photons proportional to
$g \sin\theta_{\smallw}$, we can therefore introduce an effective
electromagnetic coupling constant
\be
\alpha^{tree}_{G_{\mu}} = \frac{\sqrt{2} G_{\mu} \sdw \mw^2}{\pi}
\label{alphagmu}
\ee
which is derived from eq.~\myref{gdr} and is
evaluated in tree-level approximation by setting 
$\Delta r=0$.
The effective coupling $\alpha_{G_{\mu}}$ differs from the fine
structure constant $\alpha$, evaluated at zero momentum transfer, by
higher-order effects.
\subsubsection{Bremsstrahlung corrections}
\begin{figure}
\begin{center}
\includegraphics[scale=0.95]{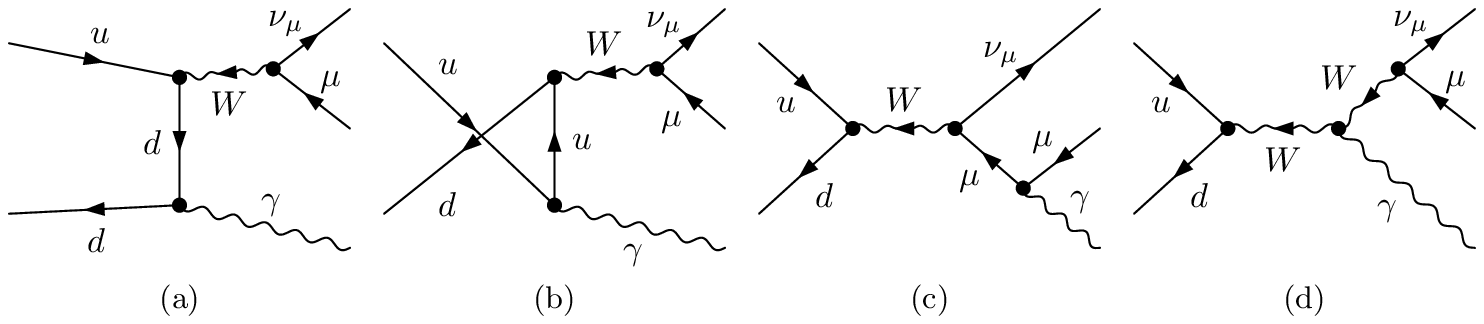}
\caption{\oa bremsstrahlung diagrams.}
\label{figreal}
\end{center}
\end{figure}

The real radiative corrections to the charged Drell-Yan process,
described by the amplitude ${\cal M}_{1}$,
are given by all the Feynman diagrams (figure~\ref{figreal})
with the emission of one extra
photon, from all the electrically charged legs of the Born diagram, 
including the internal $W$ boson.

The probability amplitude has been calculated in the unitary gauge
with massive fermions.
We integrate the squared matrix element 
over the whole photon phase space and split the allowed
photon energy range in two intervals,
$[\lambda,\Delta E]$ and $[\Delta E,  E_{max}]$. 
The cut-off $\Delta E\ll \sqrt{s}$ is chosen in such a way that the photon
with smaller energy is considered soft and does
not modify the $2\to 2$ kinematics of the Born amplitude.
The small photon mass $\lambda$ has been introduced to regularize
the infrared divergence.
In this energy region the phase space integral,
including the full angular integration, can be solved analytically.
The result can be expressed in a factorized form, as
\be
\int_{\Omega} \frac{d^3{\mathbf k}_{\gamma}}{(2\pi)^3 2 E_{\gamma} }
|{\cal M}_{1}|^2 
~=~
|{\cal M}_0|^2~~\sum_{f=u,{\bar d},e^+}~\delta_{SB}(f,\lambda)
\ee
where the soft Bremsstrahlung factor, see e.g. \cite{Denner},
depends on the mass and electric charge of the external
radiating particles
and the phase-space region $\Omega$ is defined by the request that
the photon energy $E_{\gamma}$ satisfies
$\lambda\leq E_{\gamma} \leq \Delta E$.
We have explicitly checked that the sum of the virtual and soft-real
contributions is independent of the choice of the photon mass
$\lambda$, in the limit of small $\lambda$ values.

In the hard energy region the phase space integration has been
performed numerically, with Monte Carlo techniques
improved by importance sampling to take care of collinear and infrared
singularities, as well as the peaking behaviour around the $W$ resonance.
The sum of the soft and of the hard photon cross sections
 is independent of the cut-off $\Delta E$.
We have checked the independence of our numerical results from the
choice of the infrared separator $\varepsilon\equiv \Delta E/E$
for $10^{-8} \leq \varepsilon \leq 10^{-4}$.

The squared matrix element $|{\cal M}_{1}|^2$ has been
calculated and evaluated with good agreement in several different ways,
using $\tt FeynArts$ and $\tt FormCalc$, by hand with the help of 
$\tt FORM$~\cite{form}, using the $\tt ALPHA$ algorithm~\cite{ALPHA},
in order to check the numerical stability of the results.

\section{Higher-order corrections and matching procedure}
\label{matching}
In this section we describe the matching of the
fixed EW \oa calculation with
higher-order QED corrections (cfr. ref. \cite{BCMNP}).
The latter can be included in a generic scattering cross section
in the QED Parton Shower approach, which resums to all orders the
leading logarithmic effects.
At \oa the Parton Shower reproduces only the QED leading-log
approximation of the exact EW \oa calculation, presented in section
\ref{oa}.
We would like to combine the exact \oa results 
and QED higher orders,
to improve the approximation intrinsic to the Parton Shower,
avoiding at the same time double countings at \oab.

The matching of the two calculations is a non trivial task and we
present, for the sake of clarity,
the main ideas in a simplified unphysical toy model, namely in the case
of a scattering process where only one external particle can radiate.
A general expression of its cross section with the emission of an
arbitrary number of photons, in leading-log approximation,
can be cast in the following way:
\be
d\sigma^{\infty}=
\Pi(Q^2,\varepsilon)~
\sum_{n=0}^\infty \frac{1}{n!}~|{\cal M}_{n,LL}|^2~d\Phi_n
\label{general}
\ee
where $\Pi(Q^2,\varepsilon)$ is the Sudakov form-factor that accounts for
soft-photon (up to $\varepsilon$) and virtual emissions and $Q^2$ is
related to the energy scale of the hard scattering process.
$|{\cal M}_{n,LL}|^2$ is the squared amplitude in leading-log
approximation describing the process with
the emission of $n$ hard photons, with energy larger than
$\varepsilon$ in units of the radiating particle energy. 
$d\Phi_n$ is the exact phase-space element of the
process, with $n$ photons in the final state, divided by the incoming
flux factor.
The cross section $d\sigma^{\infty}$ is independent of the infrared
separator $\varepsilon$.

According to the theorems for the factorization of soft and collinear
photon emission, the leading part of the squared amplitudes can be
written  in a factorized form;
for example the one-photon emission squared amplitude reads~\cite{PSC}
\be
|{\cal M}_{1,LL}|^2~=~
\frac{\alpha}{2\pi}~
P(z)~
I(k)~
\frac{8\pi^2}{E^2 z (1-z)}~
|{\cal M}_0|^2
\label{onegamma}
\ee
where $1-z$ is the fraction of energy carried by the photon,
$I(k)$ is a function which describes the angular spectrum of the photon
and $P(z)=(1+z^2)/(1-z)$ is the Altarelli-Parisi 
$e \to e+\gamma$ splitting function. 
In eq.~\myref{onegamma} we observe the factorization of the Born squared
amplitude and that the emission factor can be iterated to all orders,
giving $|{\cal M}_{n,LL}|^2$.
The Sudakov form factor $\Pi(Q^2,\varepsilon)$ can be expressed as
\be
\Pi(Q^2,\varepsilon)=
\exp\left(
-\frac{\alpha}{2\pi}~I_+~\log\frac{Q^2}{m^2}
\right),~~~~~~
I_+\equiv
\int_0^{1-\varepsilon}
dz P(z)
\ee
The function $I(k)$ has the property that
$
\int d\Omega_{\gamma}~I(k)=\log Q^2/m^2
$
and allows the cancellation of the infrared logarithms.

It would be desirable to include in eq.~\myref{general}
the missing \oa contributions.
The matching procedure can be better understood by comparing the exact
\oa cross section with the \oa expansion of eq.~\myref{general},
which, as we already mentioned,
does not coincide, by definition, with an exact \oa result.
In fact
\bea
d\sigma_{\alpha,LL}&=&
\left[
1-\frac{\alpha}{2\pi}~I_+~\log\frac{Q^2}{m^2}
\right] |{\cal M}_0|^2 d\Phi_0
+
|{\cal M}_{1,LL} |^2 d\Phi_1\nonumber\\
&\equiv&
\left[
1+C_{\alpha,LL}
\right] |{\cal M}_0|^2 d\Phi_0
+
|{\cal M}_{1,LL} |^2 d\Phi_1
\label{LL1}
\eea
whereas an exact NLO cross section can be always cast in the form
\be
d\sigma_{\alpha}
=
\left[
1+C_{\alpha}
\right] |{\cal M}_0|^2 d\Phi_0
+
|{\cal M}_{1} |^2 d\Phi_1
\label{exact1}
\ee
The coefficient $C_{\alpha}$ contains the complete virtual \oa
and the \oa soft-bremsstrahlung squared matrix elements, in units of
the Born squared amplitude
and $|{\cal M}_1|^2$ is the exact squared matrix element with the
emission of one hard photon.
We remark that $C_{\alpha,LL}$ and $|{\cal M}_{1,LL}|^2$
have the same singular logarithmic structure of
$C_{\alpha}$ and of $|{\cal M}_1|^2$.

We observe that, by introducing the correction factors
\be
F_{SV}~=~
1+\left(C_\alpha-C_{\alpha,LL}\right),~~~~~~
F_H~=~
1+\frac{|{\cal M}_1|^2-|{\cal M}_{1,LL}|^2}{|{\cal M}_{1,LL}|^2},
\label{FSVH}
\ee
the exact \oa cross section can be expressed, up to terms of
${\cal O}(\alpha^2)$,
in terms of its leading-log approximation as
\be
d\sigma_\alpha~=~
F_{SV} (1+C_{\alpha,LL} ) |{\cal M}_0|^2 d\Phi_0
~+~
F_H |{\cal M}_{1,LL}|^2 d\Phi_1
\label{matchedalpha}
\ee
Driven by eq.~\myref{matchedalpha},
we can improve eq.~\myref{general}
by writing the resummed cross section as
\be
d\sigma^{\infty}=
F_{SV}~\Pi(Q^2,\varepsilon)~
\sum_{n=0}^\infty \frac{1}{n!}~
\left( \prod_{i=0}^n F_{H,i}\right)~
|{\cal M}_{n,LL}|^2~
d\Phi_n
\label{matchedinfty}
\ee
The correction factors $F_{H,i}$ follow from the definition eq.~\myref{FSVH}
for each photon emission.
The expansion at \oa of eq.~\myref{matchedinfty} coincides now with
the exact NLO cross section of eq.~\myref{exact1}.
Furthermore, all higher-order leading-log contributions
are the same as in eq.~\myref{general}.
It is worth noticing that $F_{SV},F_{H,i}$ are, by construction,
infrared safe and free of collinear logarithms.

Alternatively, we remark that
one could have improved eq.~\myref{general} by adding the
\oa\, contributions missing in the leading-log approximation.
However, we prefer the factorized formulation eq.~\myref{matchedinfty}
for different reasons.
The first is that it allows to include a large class of radiative
corrections to all orders beyond the leading logarithmic approximation.
In fact, the factorization of the soft and collinear QED corrections 
is a universal property valid for any hard scattering process;
the latter can be computed not only at tree level, 
but also including infrared safe radiative corrections;
this is the case in eq.~\myref{matchedinfty}, with the factors
$F_{SV}$ and $F_{H,i}$.
As a consequence, not only the tower of leading terms of the form
$\alpha^n\log^n(s/M^2)$
but also all the terms of the form
$\alpha^n\log^{n-1}(s/M^2)\log(s/\Delta E^2)$ ($M$ is the mass of the
radiating particle)
are correctly taken into account by eq.~\myref{matchedinfty}.
Another reason for choosing a factorized formulation
is related to the request that the exponentiated
cross section should go to zero in the limit of vanishing infrared
separator $\varepsilon$: in the addittive formulation the missing
infrared finite \oa terms would spoil this property.

Equation~\myref{matchedinfty} is our master formula for the matching between
the exact \oa EW calculation and the QED resummed Parton Shower
cross section.
Its extension to the
realistic case, where every charged particle radiates photons,
is almost straightforward.

It is useful to present, in the general case, the expression of the
function $I(k)$, which describes the leading behaviour of the angular spectrum
of the emitted photons.
\be
I(k)~=~
\sum_{i,j}~
Q_i Q_j \eta_i \eta_j~
\frac{p_i\cdot p_j}{p_i\cdot k~~p_j\cdot k}~
E_\gamma^2
\ee
where $Q_l$ and $p_l$ are the electric charge fraction 
and the momentum of the external fermion $l$, 
$\eta_l$ is a charge factor equal to +1
for incoming particles or outgoing antiparticles
and equal to -1 for incoming antiparticles or outgoing particles,
$k$ is the photon momentum
and the sum runs over all the external fermions.

In our case the Sudakov form factor is
\be
\Pi(p_i;\varepsilon)
~=~
\exp
\left(
-\frac{\alpha}{2\pi}
I_+~
\int d\Omega_\gamma~I(k)
\right)
\ee
The result of the analytical integration of $I(k)$ is:
\be
\int d\Omega_\gamma I(k) =
-\sum_{j>i}\sum_i\eta_i\eta_jQ_iQ_j\frac{2 \; \tilde{p}_i\cdot
p_j}{\tilde{p}^2_i - p^2_j} \ \log\frac{\tilde{p}^2_i}{p^2_j} -
\sum_i Q^2_i
\ee
where
$$ \tilde{p}_i = \beta_{ij} p_i$$
The parameters $\beta_{ij}$ are obtained requiring that
$(\tilde{p}_i-p_j)^2 = 0$ and $\tilde{p}_i^0-p_j^0 = \beta_{ij}p_i^0 -
p_j^0 > 0$.
\section{Hadron-level cross section}
\label{hadronic}
In this section we discuss the hadron-level cross section 
$\sigma(p\smartpap\to l\nu_l +n\gamma +X) $
and the procedure to subtract the initial-state mass
singularities in the calculation of the cross section first with
\oa and then with the \oa improved with QED higher-order corrections. This
procedure makes the resulting hadron-level cross section independent of
the (unphysical) value of the initial-state quark masses.

The radiative corrections are enhanced by large collinear logarithmic
(mass singularities) factors,
of the form $\alpha/(2\pi) \log(s/m_f^2)$.
The initial-state collinear logarithms 
are universal, i.e. are independent of the hard scattering process, 
must be factorized out at an energy scale $M$
and can be reabsorbed in the definition of the parton densities which
describe the partonic content of the proton, like in all NLO QCD calculations.
Since the partonic cross section described in section \ref{oa} still
contains QED initial-state mass singularities,
it is mandatory to implement a subtraction procedure 
to avoid a double counting when convoluting
with the proton parton densities.
Recently a new set of PDFs ($\tt MRST2004QED$) including QED effects has been 
published~\cite{MRST2004QED}. These parton densities are evolved via
the DGLAP equations including also the QED splitting functions.
As discussed in the literature \cite{SP,RW},
the effect of the QED evolution on the PDFs is at the per mille level,
for $x$ and $M^2$ values probed by Drell-Yan dynamics at the Tevatron
and at the LHC.

The subtraction at \oa has been discussed for example in ref.~\cite{DK} and
is obtained by a redefinition of the parton densities.
The hadron-level cross section at \oa can be written
\bea
\label{sigmahad}
d\sigma(p\smartpap\to l\nu_l+X)&=&
\sum_{a,b}\int_0^1 dx_1 dx_2~~
q_a(x_1,M^2) q_b(x_2,M^2)
\left[
d{\sigma}_0~+~
d{\sigma}_{\alpha}
\right]-\\
&&-~\left( 
\Delta q_a(x_1,M^2) q_b(x_2,M^2)+
q_a(x_1,M^2) \Delta q_b(x_2,M^2)
\right)~
d\sigma_0
\nonumber
\eea
where $a,b$ run over all parton species described by the densities 
$q_i(x,M^2)$, $M$ is the factorization scale,
$d\sigma_0$ and $d\sigma_\alpha$ are the Born and \oa partonic
cross sections.
The \oa subtraction term is
\be
\Delta q_i(x,M^2)~=~\int_z^1 q_i\left(\frac{x}{z},M^2\right)
\frac{\alpha}{2\pi}
Q_i^2
\left[
P(z)
\left(
\log\left(\frac{M^2}{m_i^2}\right)-2 \log(1-z)-1
+f(z)
\right)
\right]_+.
\label{deltaq}
\ee
$Q_i$ and $m_i$ are the electric charge fraction and the mass of the quark $i$
and the function $f(z)$~\cite{BW} allows to change the subtraction scheme
(e.g. DIS or $\overline{MS}$).
Given the presence in the hadron-level cross section eq.~\myref{sigmahad} 
of the product of two parton densities, the subtraction procedure in a
factorized form could yield
terms of ${\cal O}(\alpha^2)$ which have been discarded for 
consistency at \oab.

It is mandatory to generalize the independence from the value of the
quark masses of the \oa cross section of eq.~\myref{sigmahad} to the
cross section including also QED higher-order corrections.
The exponentiation in the first line of eq.~\myref{sigmahad} of 
$[d\sigma_0+d\sigma_\alpha]$,
according to the matching algorithm described in section \ref{matching}, 
would develop also higher-order initial-state mass singularities, 
which appear in the form $\alpha^n\log^n(s/m_f^2) ~~(n>1)$.
In order to remove the latter in a systematic way,
we prefer to rewrite the partonic cross section in eq.~\myref{sigmahad};
we split it in two terms: one free of initial-state mass
singularities, $d\tilde\sigma_\alpha$, 
which will be then improved with the resummation, and
one which contains the singular part at \oa,~$d\sigma_\alpha^{sub}$.
We rewrite eq.~\myref{sigmahad} by adding and subtracting the same
quantity, namely
\be
d\sigma_0+d\sigma_\alpha~\to~d\sigma_0+(d\sigma_\alpha-d\sigma_\alpha^{sub})
~+~d\sigma_\alpha^{sub}
\equiv 
d\sigma_0+d\tilde\sigma_\alpha~+~d\sigma_\alpha^{sub}
\ee
The subtraction term is further split into a soft+virtual and a hard
photon contributions:
\bea
\label{subtraction}
d\sigma_\alpha^{sub}
&\equiv&
d\sigma_\alpha^{SV,sub}~+~d\sigma_\alpha^{H,sub}\\
d\sigma_\alpha^{SV,sub} &=&
-\frac{\alpha}{2\pi}
\left[
Q_a^2 \left(\log\frac{M^2}{m_a^2}-1\right)+
Q_b^2 \left(\log\frac{M^2}{m_b^2}-1\right)
\right] \cdot
I_+
|{\cal M}_0|^2 d\Phi_0
\equiv~
C_\alpha^{sub}~|{\cal M}_0|^2 d\Phi_0
\nonumber\\
d\sigma_\alpha^{H,sub} &=&
\frac{8\pi \alpha}{E^2 z (1-z)}~
\frac{1+z^2}{1-z}~
|{\cal M}_0(s^\prime)|^2~
I_{sub}(k)~d\Phi_1
~\equiv~
|{\cal M}_{1,sub}|^2~d\Phi_1
\nonumber\\
I_{sub}(k)&=&
\sum_{i=a,b}
Q_i^2
\left(
\frac{1}{1-\beta_i c} - \frac{2 m_i^2}{M^2 (1-\beta_i c)^2}
\right)\nonumber
\eea
where the subtraction cross sections are defined in analogy to the
leading log cross sections of eq.~\myref{LL1}
and $I_{sub}(k)$ describes the initial-state radiation.
In eq.~\myref{subtraction}
$|{\cal M}_0(s^\prime)|^2$ is the Born squared amplitude evaluated at
a reduced center-of-mass energy $s^\prime$,
$\beta_i=\sqrt{1-4 m_i^2/M^2}$
and $c$ is the
cosine of the angle of the photon with the beam axis in the 
partonic center-of-mass frame.
Once integrated over the photonic variables,
$d\sigma_\alpha^{H,sub}$ develops the same infrared
logarithmic structure as $d\sigma_\alpha^{SV,sub}$ and the sum is
independent of the value chosen for the infrared separator.

By applying the matching algorithm to
$[d\sigma_0+d\tilde\sigma_\alpha]$, which is free of collinear
initial-state logarithms,
we obtain an improved hadron-level cross section including QED
higher-order corrections.
Starting from the \oa expression, identical to eq.~\myref{sigmahad},
\bea
\label{sigmahadsubalpha}
d\sigma_{had}&=&
\sum_{a,b}\int_0^1 dx_1 dx_2~~
q_a(x_1,M^2) q_b(x_2,M^2)
\left[
d{\sigma}_0~+~
d{\tilde\sigma}_{\alpha}
\right]+\\
&+&q_a(x_1,M^2) q_b(x_2,M^2)
\left[
d\sigma_\alpha^{sub}
-
\left(
 \frac{\Delta q_a(x_1,M^2)}{q_a(x_1,M^2)}
+\frac{\Delta q_b(x_2,M^2)}{q_b(x_2,M^2)}
\right)
d\sigma_0
\right]
\nonumber
\eea
we obtain the hadron-level QED resummed cross section
\bea
\label{sigmahadsubinfty}
d\sigma_{had}&=&
\sum_{a,b}\int_0^1 dx_1 dx_2~~
q_a(x_1,M^2) q_b(x_2,M^2)\times\\
&\Bigg\{&
{\tilde F}_{SV}~{\tilde\Pi}(Q^2,\varepsilon)~
\sum_{n=0}^\infty \frac{1}{n!}~
\left( \prod_{i=0}^n {\tilde F}_{H,i}\right)~
|{\tilde{\cal M}}_{n,LL}|^2~
d\Phi_n~+
\nonumber\\
&&~~+
\left[
d\sigma_\alpha^{sub}
-
\left(
 \frac{\Delta q_a(x_1,M^2)}{q_a(x_1,M^2)}
+\frac{\Delta q_b(x_2,M^2)}{q_b(x_2,M^2)}
\right)
d\sigma_0
\right]
\Bigg\}.\nonumber
\eea
The variables with a tilde represent quantities subtracted of the
initial-state singularities; more precisely
the matching of section \ref{matching} is built
using the following subtracted quantities:
\bea
&&|{\tilde{\cal M}}_{1}|^2~=~|{\cal M}_1|^2~-~|{\cal M}_{1,sub}|^2,~~~~~~
|{\tilde{\cal M}}_{1,LL}|^2~=~|{\cal M}_{1,LL}|^2~-~|{\cal  M}_{1,sub}|^2
\nonumber\\
&&{\tilde C}_\alpha~=~C_\alpha-C_\alpha^{sub},~~~~~~~
{\tilde C}_{\alpha,LL}~=~C_{\alpha,LL}-C_\alpha^{sub},~~~~~~~
{\tilde I}(k) ~=~ I(k)~-~I_{sub}(k)\nonumber
\eea
The \oa expansion of eq.~\myref{sigmahadsubinfty} coincides with
eq.~\myref{sigmahadsubalpha} or equivalently to eq.~\myref{sigmahad}.
The second line in eq.~\myref{sigmahadsubinfty} is by construction free
of initial-state mass singularities and we remark that the same
property holds also for the finite \oa correction of the last line.

Equation~\myref{sigmahadsubinfty} is our master formula for the computation of
the hadron-level cross sections and the event simulation.
\section{Numerical results}
\label{results}
All the numerical results have been obtained using the following
values for the input parameters:
\begin{center}
\begin{tabular}{lll}
$\alpha=1/137.03599911$ & 
$G_{\mu} = 1.16637~10^{-5}$ GeV$^{-2}$ &
$\mz=91.1876$~GeV \\
$\mw = 80.425$~GeV & 
$\gw = 2.124$~GeV &
$\mh = 115$~GeV\\
$m_e=510.99892$~KeV &
$m_{\mu}=105.658369$~MeV &
$m_{\tau}=1.77699$~GeV \\
$m_u = 66$~MeV &
$m_c = 1.2$~GeV &
$m_t = 178$~GeV \\
$m_d = 66$~MeV &
$m_s = 150$~MeV &
$m_b = 4.3$~MeV \\
$V_{ud}=0.975$ &
$V_{us}=0.222$ &
$V_{ub}=0$ \\
$V_{cd}=0.222$ &
$V_{cs}=0.975$ &
$V_{cb}=0$ \\
$V_{td}=0$ &
$V_{ts}=0$ &
$V_{tb}=1$ \\
\end{tabular}
\end{center}
and have been computed
in the $G_{\mu}$ input scheme described in Section~\ref{oa}. 
The set of parton density functions used to compute all the hadron-level
cross sections is 
$\tt MRST2004QED$~\cite{MRST2004QED}~\footnote{Since this set of PDFs 
includes a photon distribution function resulting from the QED evolution
of the PDFs, the Drell-Yan cross section receives a new type of 
correction from $2 \to 3$ photon-induced processes, such as, for example, 
$\gamma q \to q^{\prime} l \nu_l$.
These contributions, which have been evaluated 
by Dittmaier and Kr\"amer in ref. \cite{LesHouches},
are not considered in the present study.}.
In this set of PDFs the QCD and the QED factorization scales are set
to be equal and,
as usually done in the literature \cite{DK, BW}, 
we choose $M=\mw$, if not stated otherwise.
The use of the PDFs set $\tt MRST2004QED$
implies that our numerical results  correspond to the DIS
factorization scheme.
The computation of the hadron-level results requires the numerical evaluation
of the subtraction term defined in eq. \ref{deltaq};
a grid of values in the variable $x$, which is then interpolated,
is obtained by means of the numerical library $\tt CUBA$ \cite{CUBA}.
All the hadron-level results refer to the LHC, at a nominal
center-of-mass energy $\sqrt{s}=14$ TeV, if not stated otherwise.

The fine structure constant $\alpha$ is used
instead of $\alpha^{tree}_{G_\mu}$ of eq.~\myref{alphagmu}
in the computation of
all the \oa and higher-order corrections,
in order to describe the real photon emission with the proper coupling.
This replacement is formally justified
in the fixed \oa calculation because it differs at ${\cal O}(\alpha^2)$.
The $G_\mu$ input scheme has been implemented by computing
all the contributions to the cross section with $\alpha=\alpha(0)$,
subtracting the $\Delta r$ contribution, evaluated as well with $\alpha(0)$,
and then rescaling the total result by $(\alpha^{tree}_{G_\mu}/\alpha(0))^2$.

The following cuts have been imposed to select the events:
\be
p_{\perp,\ell}>25~\mbox{GeV},\qquad
p_{\perp,missing}>25~\mbox{GeV},\qquad
|\eta_{\ell}|<2.5
\label{cuts}
\ee
where $p_{\perp,\ell}$ and $\eta_\ell$ are the transverse momentum and
the pseudo-rapidity of the charged lepton and
$p_{\perp,missing}$ is the missing transverse momentum, which in our
case coincides with the neutrino $p_\perp$.

Our results are obtained with bare muons and with ``recombined''
electrons.
We assume perfect isolation of photons from the muon, which is
experimentally achievable with good accuracy:
the resulting cross sections are therefore enhanced by large muon mass
collinear logarithms, because the photon emission is not treated
inclusively in the region about the muon.
In the case of electrons, it is experimentally difficult to separate
them from the photon track, when the latter lyes within a
cone around the lepton. We adopt the following recombination
procedure:
\begin{itemize}
\item photons with a rapidity $|\eta_\gamma|>2.5$ are
never recombined to the electron;
\item if the photon rapidity is $|\eta_\gamma|<2.5$ and
$R_{e\gamma}=\sqrt{(\eta_e-\eta_\gamma)^2+\phi_{e\gamma}^2 }<0.1$
($\phi_{e\gamma}$ is the angle between the photon and the
electron in the transverse plane), then the photon is recombined with the
electron, i.e. the momenta of the two particles are added and
associated with the momentum of the electron;
\item the resulting momenta should satisfy the cuts of
eq.~\myref{cuts}.
\end{itemize}

In order to study the different effects of the radiative corrections
on the relevant observables, we will distinguish the approximations
described in table~\ref{tableapprox}.
\begin{table}
\begin{center}
\begin{tabular}{|c|l|}
\hline
1. & lowest order (Born) \\
2. & final state \oa LL QED Parton Shower \\
3. & final state exponentiated LL QED Parton Shower \\
4. & exact \oa EW  of eq.~\myref{sigmahad}\\
5. & exact \oa EW matched with higher-order QED corrections (best) of eq.~\myref{sigmahadsubinfty}\\
\hline
\end{tabular}
\caption{Different approximations for the cross section of the
  Drell-Yan charged current process.}
\label{tableapprox}
\end{center}
\end{table}
The approximations 2. and 3. refer to the old version of $\tt
HORACE$~\cite{CMNT}, which included the final-state QED corrections
in leading-log accuracy (both at \oa and with higher orders) in a
pure Parton Shower approach.
\subsection{Technical checks}
In this section we discuss some technical checks satisfied by the new
version of $\tt HORACE$, namely the independence from the
$\varepsilon$ parameter and from the quark masses. As reference, we
also provide the (un-subtracted) cross section at parton level for
three different center-of-mass energies.

The definition of the soft- and hard-bremsstrahlung corrections
requires the introduction 
of an infrared separator $\varepsilon$.
The physical cross section does not depend on
this parameter as shown in table~\ref{tabesoftmax}.
\begin{table}
\begin{center}
\begin{tabular}{|c|c|c|}
\hline
$\varepsilon$ & \oa & best\\
\hline
$5\cdot 10^{-4}$ & $4410.98 \pm 0.20$ & $4412.14 \pm 0.26$ \\
\hline
$1\cdot 10^{-6}$ & $4410.84 \pm 0.48$ & $4413.66 \pm 0.9 $    \\
\hline
\end{tabular}
\caption{Independence of the hadron-level cross section (pb)
of the value chosen for the infrared separator $\varepsilon$.}
\label{tabesoftmax}
\end{center}
\end{table}

The partonic cross section is computed using quark masses to
regularize the initial-state collinear singularities. The latter are
subtracted according to the procedure described in
Section~\ref{hadronic} when calculating the hadron-level cross section, which
has to be independent,
up to terms of order $m^2_q/\mw^2$,
of the value chosen for the quark masses. This
property is demonstrated in table~\ref{tabquarkmasses}, both for the
\oa and matched cross section.
\begin{table}
\begin{center}
\begin{tabular}{|c|c|c|}
\hline
          & \oa              & best \\
\hline
$m_q$     & $4410.98 \pm 0.20$ & $4412.14 \pm 0.26$ \\
\hline
$m_q/10$  & $4410.92 \pm 0.26$ & $4411.89 \pm 0.33$ \\
\hline
$m_q/100$ & $4410.99 \pm 0.29$ & $4411.92 \pm 0.50$  \\
\hline
\end{tabular}
\caption{Indipendence of the hadron-level \oa and exponentiated
cross sections (pb) of the value chosen for the quark masses.}
\label{tabquarkmasses}
\end{center}
\end{table}

Finally, the partonic cross section has been computed, within the cuts
of eq.~\myref{cuts}, without subtraction of the initial-state collinear
divergences; the results are presented in table~\ref{partonicunsubtracted}.
\begin{table}
\begin{center}
\begin{tabular}{|c|c|c|c|}
\hline
$\sqrt{s}$ (GeV)& 60 & 80 & 100 \\
\hline
Born & $7.807 \pm 0.001 $ & $5223.0 \pm 0.5$ & $25.452 \pm 0.002$ \\
\hline
\oa  & $7.274 \pm 0.002 $ & $4817.1 \pm 0.8$ & $34.476 \pm 0.005$ \\
\hline
best & $7.288 \pm 0.002 $ & $4830.6 \pm 0.6$ & $34.208 \pm 0.008$ \\
\hline
\end{tabular}
\caption{Unsubtracted partonic cross section (pb) at different partonic
center-of-mass energies.}
\label{partonicunsubtracted}
\end{center}
\end{table}
\subsection{$W$ transverse mass distribution}
We present in this section the numerical results for 
the distribution of $W$ transverse mass, defined as
\be
M_\perp = \sqrt{2p_{\perp,\ell}~p_{\perp,\nu}~(1-\cos\phi_{\ell,\nu})}
\ee
where $\phi_{\ell,\nu}$ is the angle between the lepton and the
neutrino in the transverse plane.
We discuss in detail the
effect of the different classes of radiative corrections.
Their impact, relative to the Born approximation,
is in general of several per mille and in some cases of few per cent
and can play a significant role 
for instance in the precise determination of the
mass of the $W$ boson $\mw$, foreseen at the LHC with an accuracy of
15 MeV.

In table \ref{mtminup} we show the cross section integrated
within the cuts eq.~\myref{cuts} and a futher cut on the minimum $W$
transverse mass, varying from $50$ to $2000$ GeV.
The relative effect of the \oa corrections grows
with the cut on the minimum transverse mass, because of the increasing
importance of the EW Sudakov logs.
%
\begin{table}
\begin{center}
\begin{tabular}{|c|c|c|c|c|c|c|}
\hline
$m_{\perp,min}$ (GeV) & Born (pb) &  $\delta_\alpha^{\mu^+}$ (\%) & $\delta_\infty^{\mu^+}$ (\%) & $\delta_\alpha^{e^+}$ (\%)& $\delta_\infty^{e^+}$ (\%)\\
\hline
50  & $4536.03(7)$         & -2.8                  & -2.7                   & -1.7                   & -1.8  \\ 
\hline
100 & $27.642(1)$       & -5.0                  & -4.9                   & -3.4                   & -3.4  \\
\hline
200 & $1.79275(5)$     & -7.9                  & -7.7                   & -6.3                   & -6.3  \\
\hline
500 & $0.084809(2)$    & -14.3                 & -13.8                  & -12.2                  & -12.2 \\
\hline
1000& $0.0065320(2)$   & -21.9                 & -21.1                  & -19.4                  & -19.1 \\
\hline
2000& $0.000273686(8)$ & -32.1                 & -30.5                  & -28.7                  & -28.1 \\
\hline
%
\end{tabular}
\caption{Lowest-order hadron-level cross section, integrated imposing a cut 
on the
minimum transverse mass and relative effects, with respect to the Born
cross section, in the \oa
($\delta_\alpha^\ell$) and in the best ($\delta_\infty^\ell$)
approximations.} 
\label{mtminup}
\end{center}
\end{table}

The Born results coincide for muons and electrons, up to negligible
mass effects.
The radiative corrections instead differ because of the final-state collinear 
logarithmic enhancement, which are absent in the case of
photons recombined with the electron.
All the QED higher-order corrections do not modify significantly the
\oa corrections.

\begin{figure}
\begin{center}
\includegraphics[height=80mm,angle=0]{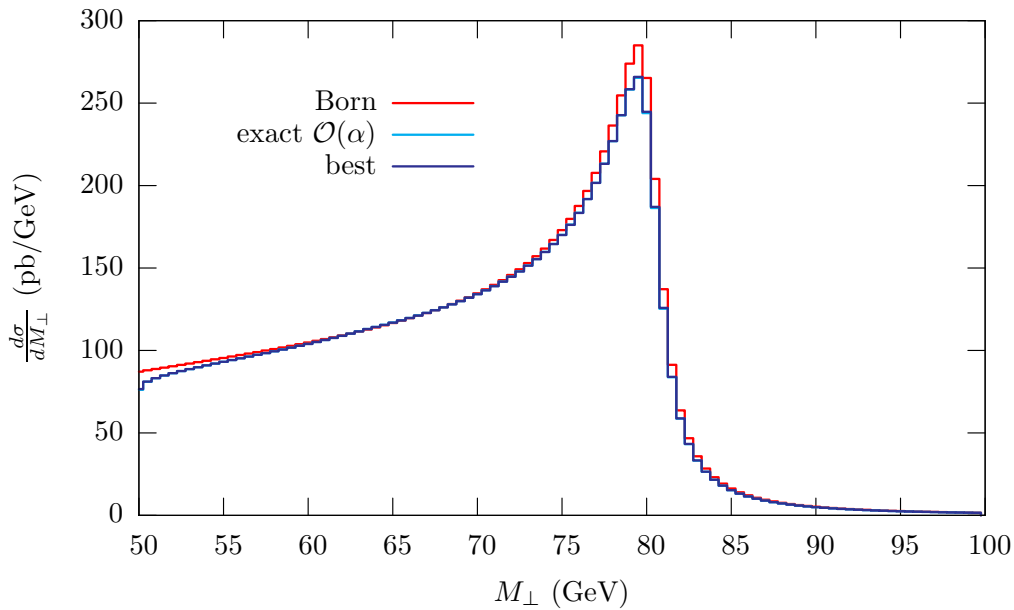}
\end{center}
\caption{Transverse mass distribution in Born,~\oa and best approximations.}
\label{mtdistr1}
\end{figure}

\begin{figure}
\begin{center}
\includegraphics[height=80mm,angle=0]{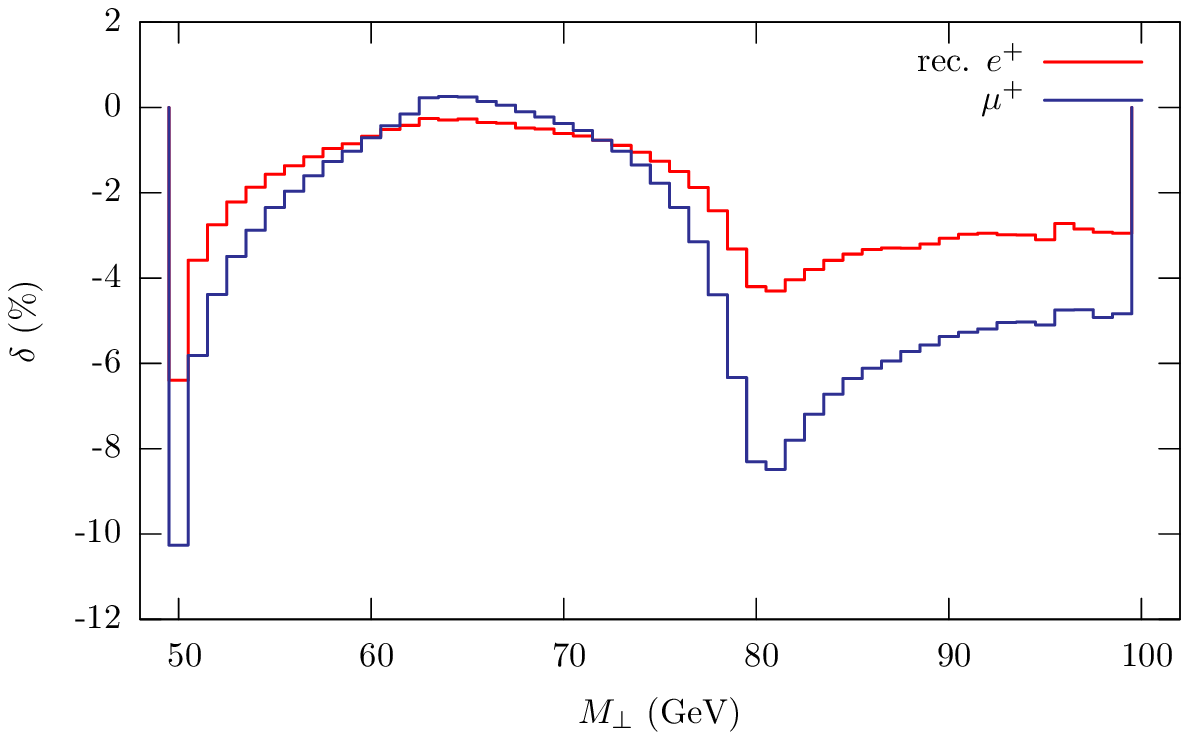}
\end{center}
\caption{Relative corrections with respect to the Born cross section due
to the exact \oa corrections for muons and recombined electrons final
states.}
\label{mtdistr2}
\end{figure}

\begin{figure}
\begin{center}
\includegraphics[height=80mm,angle=0]{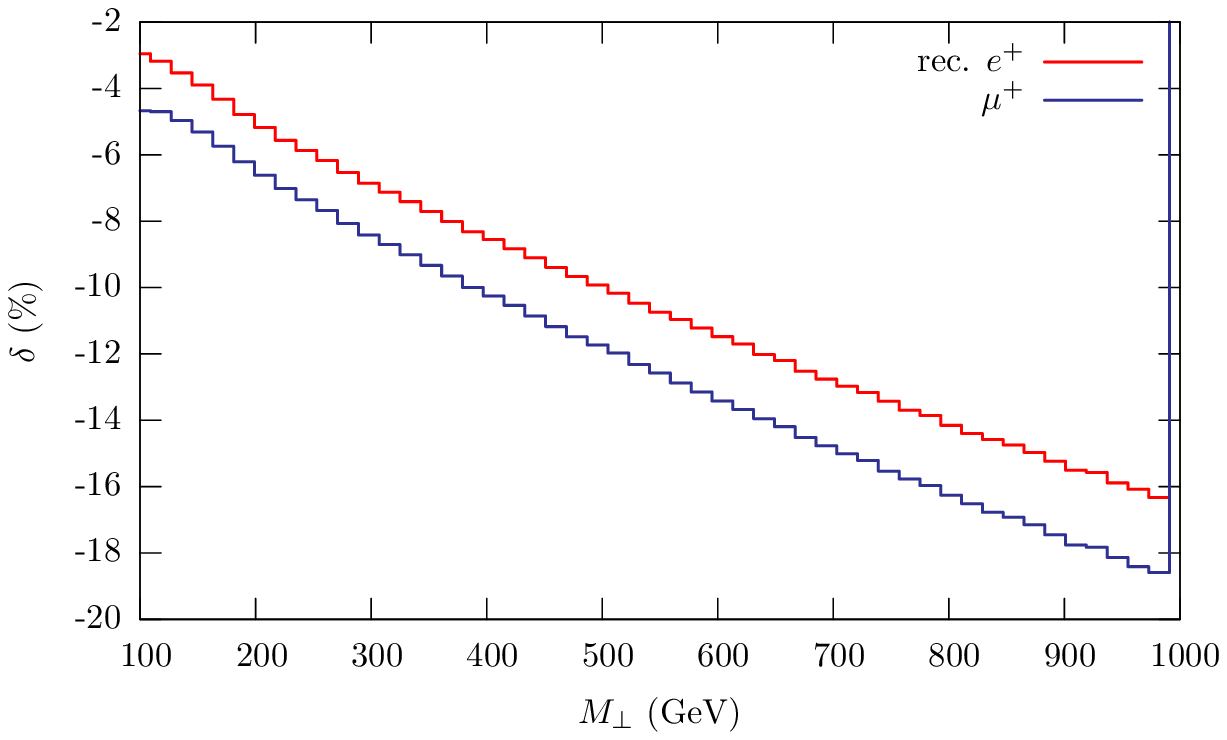}
\end{center}
\caption{Relative corrections with respect to the Born cross section due
to the exact \oa corrections for muons and recombined electrons final
states.}
\label{mtdistr3}
\end{figure}

\begin{figure}
\begin{center}
\includegraphics[height=45mm,angle=0]{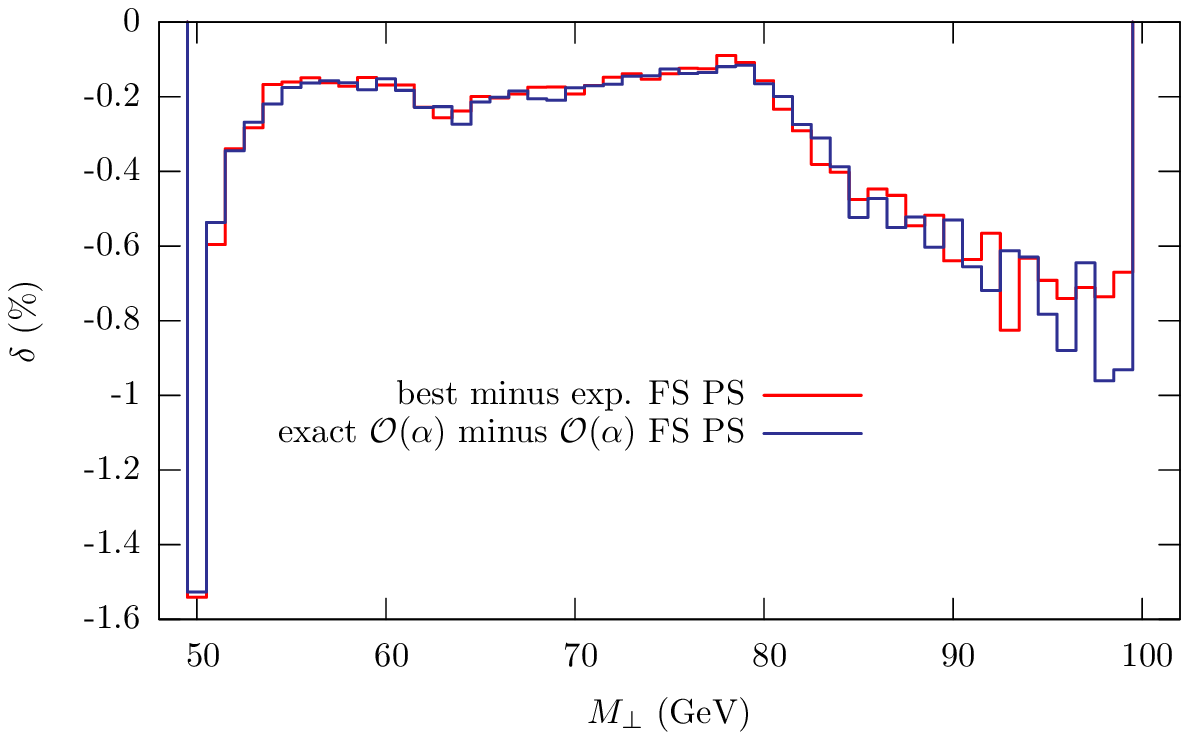}~
\includegraphics[height=45mm,angle=0]{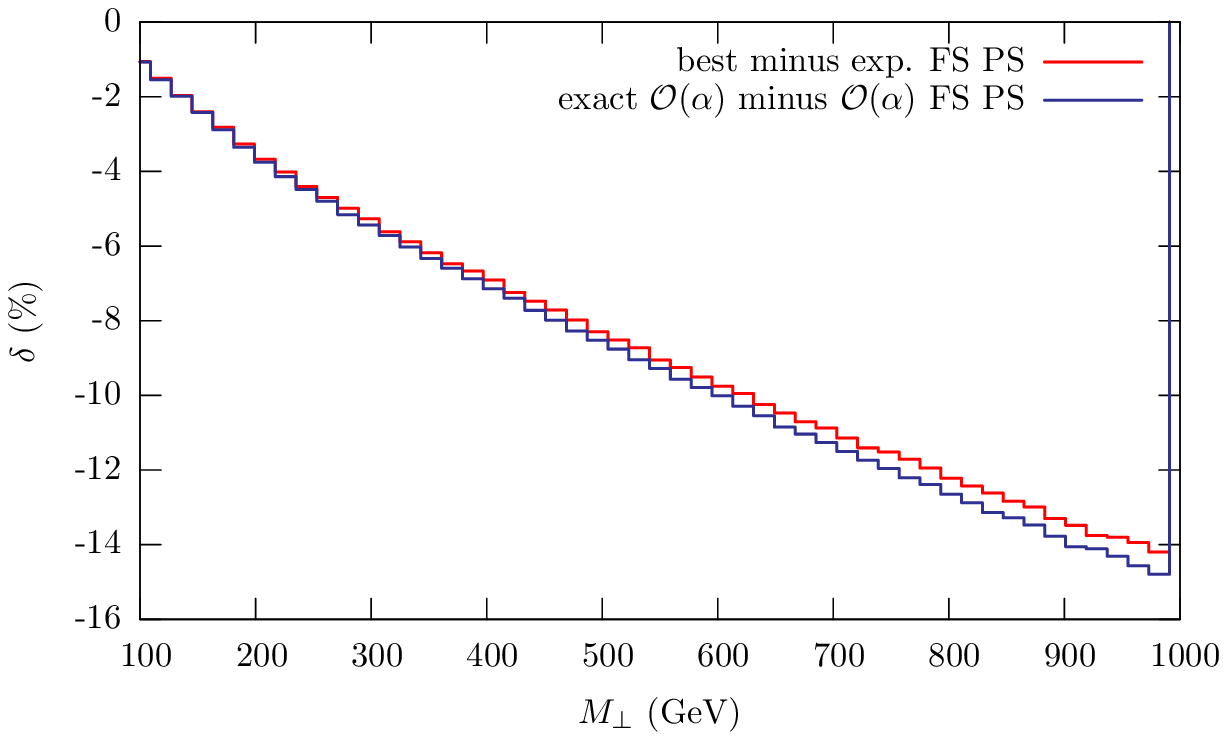}
\end{center}
\caption{Relative effect,
in Born units, of the difference between the approximations 4. and 2.
of table~1 (blue line) and between 5. and 3. (red line).
}
\label{mtdistr4}
\end{figure}

\begin{figure}
\begin{center}
\includegraphics[height=45mm,angle=0]{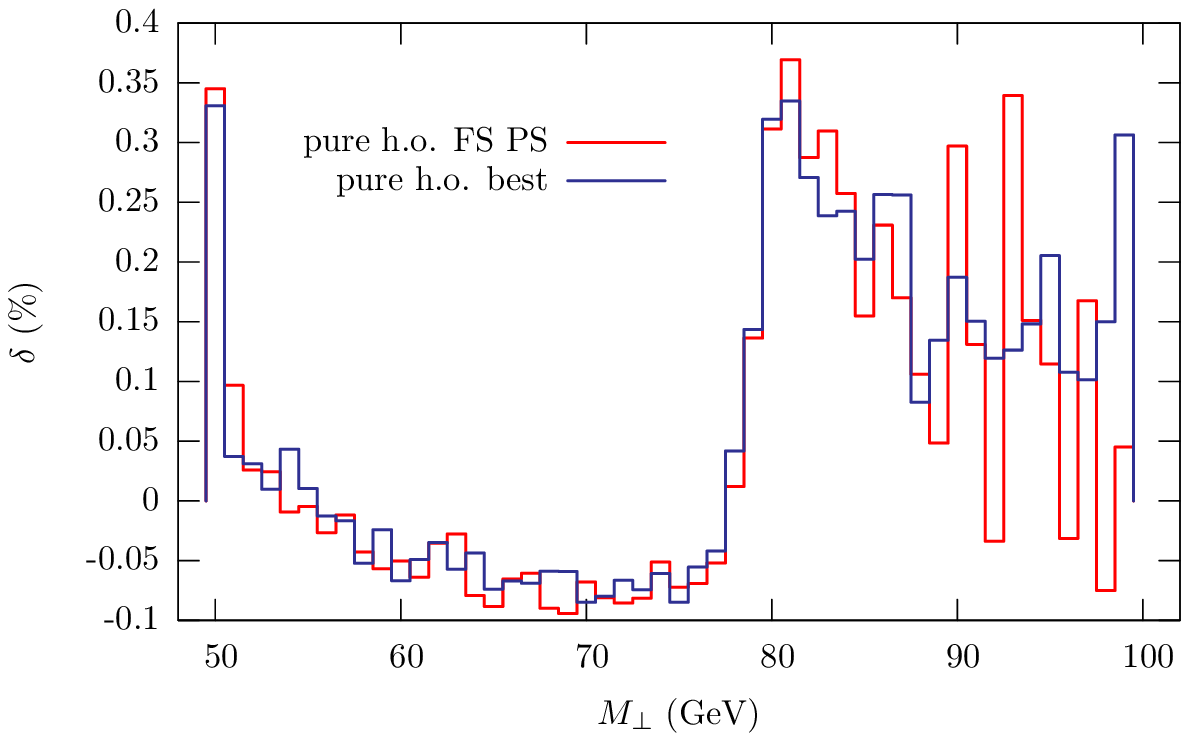}~\includegraphics[height=45mm,angle=0]{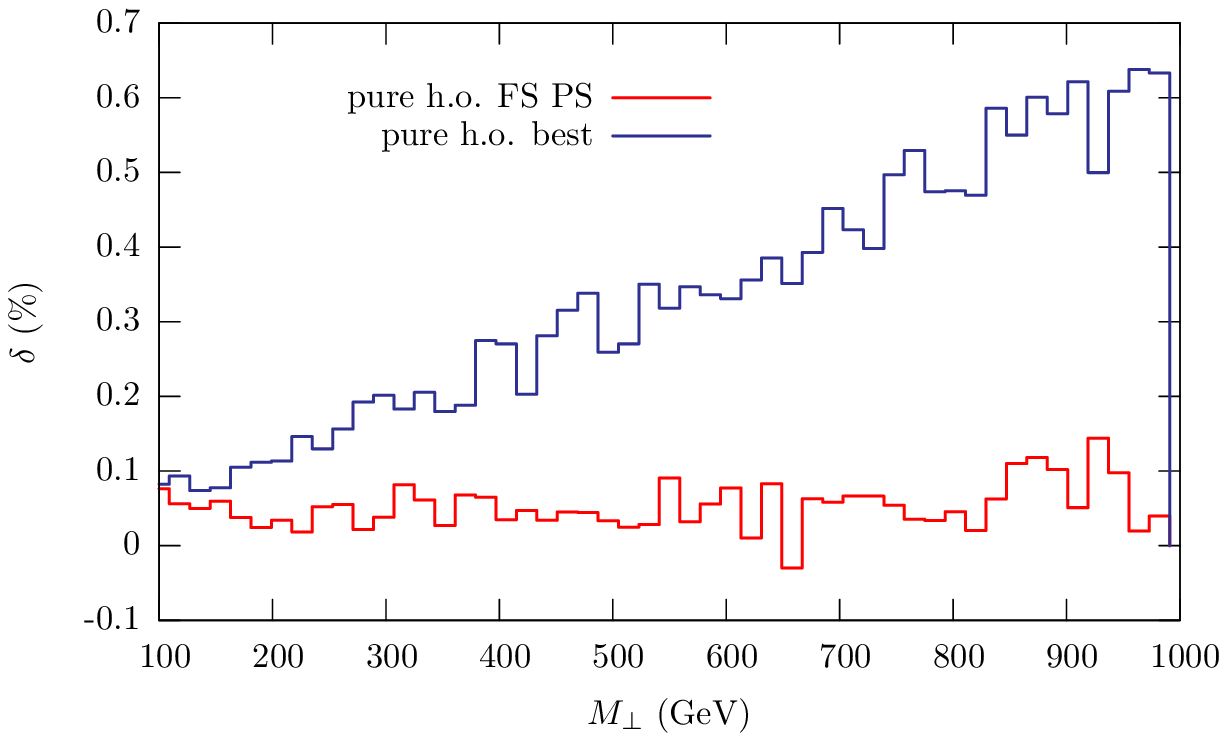}
\end{center}
\caption{Relative effect  on the transverse mass distribution, 
in Born units, of higher-order QED final state-like and
full QED parton shower corrections.}
\label{mtdistr5}
\end{figure}

In figures from \ref{mtdistr1} to \ref{mtdistr5}
we show the transverse mass distribution and
disentangle the different contributions due to the radiative corrections.
In figure~\ref{mtdistr1}
the transverse mass distribution is plotted,
in the range $50<M_\perp<100$ GeV.
The transverse mass distribution provides physical
information in different ranges:
the position of the jacobian peak and the shape of the distribution
about the peak can be used to extract the value of the $W$ boson mass,
the shape of the tail of the distribution above the peak,
$80<M_\perp<100$ GeV, can be used to measure the 
$W$ boson decay width and
the large transverse mass tail, $200<M_\perp<1000$ GeV,
of the distribution can be an important background to the
searches of new heavy gauge bosons.

In figure \ref{mtdistr2} we plot, 
in the range $50<M_\perp<100$~GeV
the effect of the exact \oa radiative correction, relative to the Born
cross section, 
in the case of muons and of recombined electrons.
The \oa contribution gives a large correction, up to $\sim-10\%$,
which distorts, about the $W$ resonance, the transverse mass distribution
and is responsible for the bulk of the shift in the extraction of the
$W$ boson mass.
As shown in figure \ref{mtdistr3}, in the range
$100<M_\perp<1000$~GeV
the EW Sudakov logarithms make the effect of the radiative corrections large and negative, reaching the 20\% level.

In figure \ref{mtdistr4}
we disentangle, among the \oa contributions, the effect of all the
corrections which can not be classified as QED final state-like
leading-log radiation, by taking (blue line) the difference between 
approximations 4. and 2. (and between 5. and 3., red line) of
table~\ref{tableapprox} in units of the differential Born cross section.
We present only
the results for muons, being the effect similar in the electron case.
We observe that they are quite flat, small and negative, for
$M_\perp<80$ GeV; they become larger in size and always
negative for increasing values of $M_\perp$,
because of the presence of the EW Sudakov logs.
From a comparison of figures \ref{mtdistr3} and \ref{mtdistr4},
the non-factorizable weak contributions account for more than half of the
\oa radiative corrections, for $M_\perp>200$ GeV.

In figure~\ref{mtdistr5}
we present the effect of the higher-order (beyond \oab) corrections,
and disentangle
the effect of all the terms which can not be classified as QED final
state-like leading log radiation, by considering the difference of the
3. and 2. (red line) and of 5. and 4. (blue line) approximations, in
units of the lowest-order differential cross section. We present only
the results for muons, being the higher-order corrections
smaller in the electron case because of the recombination.
The red line describes the effect of purely photonic final-state like
leading log corrections, whereas the blue line represents the
higher-order contributions of the matched cross section of
eq.~\myref{sigmahadsubinfty}. The latter includes, besides the content
of the red line, the remnant of the initial-state radiation after the
subtraction of the initial-state singularities and the product of purely
weak corrections (the $\tilde{F}_{SV}$ factor of
eq.~\myref{sigmahadsubinfty}) with photonic radiation.
Around the peak the two lines almost coincide, while for large
$M_\perp$ we observe the effect of the product of the EW Sudakov logs 
times the \oa photonic correction.

\begin{figure}
\begin{center}
\includegraphics[height=80mm,angle=0]{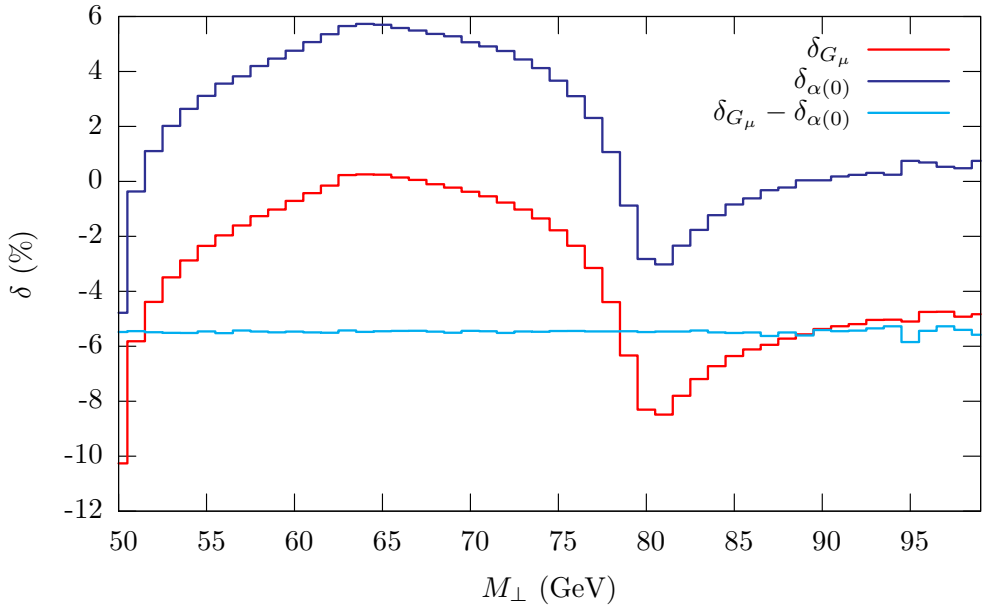}
\end{center}
\caption{Relative corrections to the transverse mass distribution
in the $G_\mu$ and in the $\alpha(0)$ schemes, expressed in units of
the corresponding Born  distributions, and their difference.}
\label{distrscheme}
\end{figure}

As we already discussed in Section~\ref{oa}, we can
compute the cross sections in the $G_\mu$ or the $\alpha(0)$ input scheme.
In table \ref{cfrschemes}, we compare the cross sections obtained 
in the two input schemes,
in Born and in \oa approximations and the corresponding relative
corrections.
The difference between the cross sections in the two schemes
is reduced when going from the Born to the \oa  approximation
and amounts to about 6\% (Born) and  1\% (\oab), respectively. 
The relative correction in the two schemes is of the same order 
($\approx 3\%$) but of opposite sign.
This can be understood taking into account
that, as previously discussed, in the $G_\mu$ scheme, at a 
variance with the $\alpha(0)$ scheme,
universal virtual corrections are absorbed in the
lowest-order cross section. 
\begin{table}
\begin{center}
\begin{tabular}{|c|c|c|c|}
\hline
scheme      & Born & \oa & $\delta$ (\%) \\
\hline
$\alpha(0)$ & $4244.68 \pm 0.09$ & $4360.5 \pm 0.6$ & +2.73  \\
\hline
$G_\mu$     & $4536.03 \pm 0.07$ & $4411.0 \pm 0.2$ & -2.76\\
\hline
\end{tabular}
\caption{Born and \oa hadron-level cross sections (pb)and effect
of the \oa corrections, expressed in units of the
corresponding Born cross section, in the
$\alpha(0)$ and in the $G_\mu$ schemes.}
\label{cfrschemes}
\end{center}
\end{table}
It is worth noticing that 
the \oa corrected transverse mass distribution differs in the two input schemes
as shown in figure \ref{distrscheme}, where we plot the relative
corrections in the two schemes in units of the corresponding Born
distributions and their difference.
\begin{figure}
\begin{center}
\includegraphics[height=75mm,angle=0]{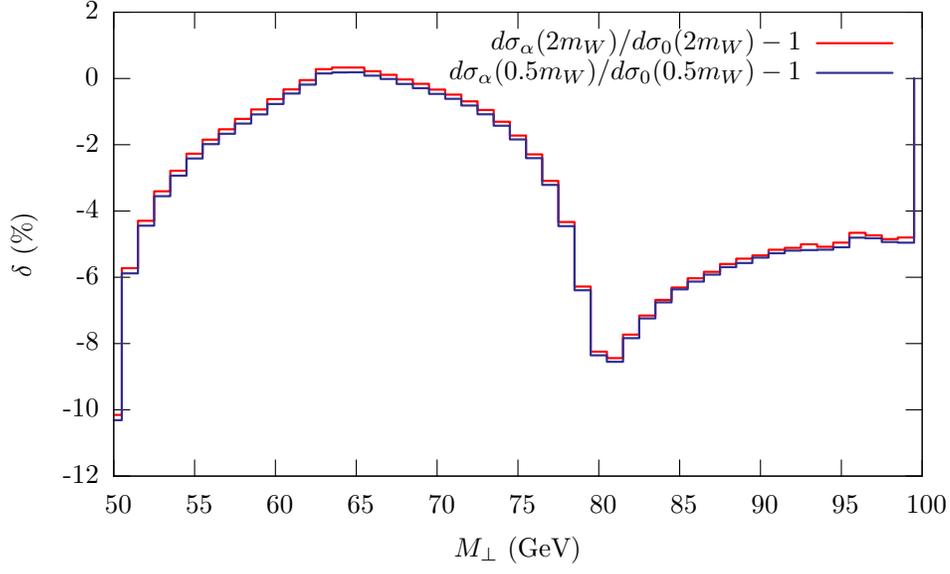}
\end{center}
\caption{Relative effect, with respect to the corresponding Born,
of the ${\cal O}(\alpha)$ corrections, 
computed with the factorization scale $M=\mw/2$ and
$M=2\mw$.}
\label{scaledep}
\end{figure}
\begin{figure}
\begin{center}
\includegraphics[height=75mm,angle=0]{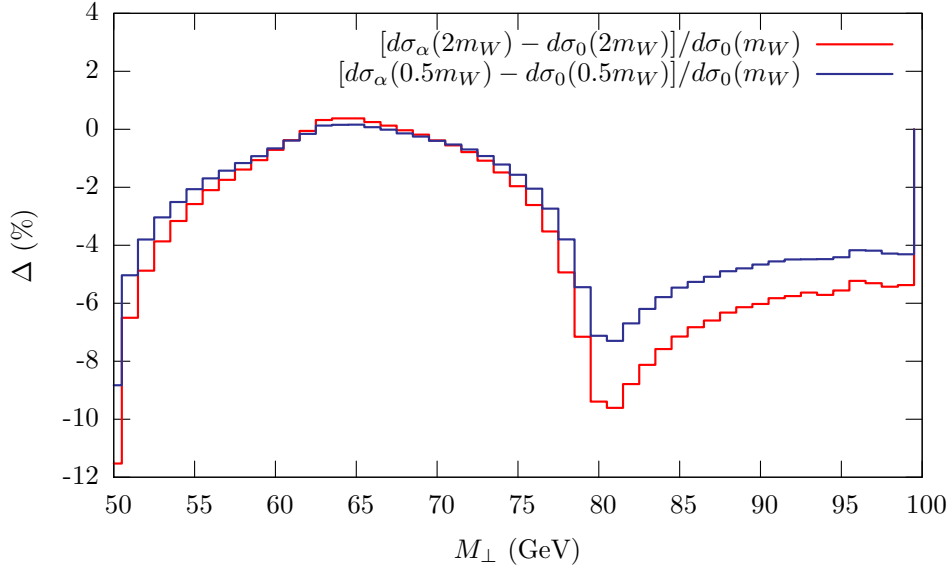}
\end{center}
\caption{Relative effect of the ${\cal O}(\alpha)$ corrections, 
computed with the factorization scale $M=\mw/2$ and
$M=2\mw$, expressed in units of the Born cross section with $M=\mw$.}
\label{scaledep2}
\end{figure}
%
%

Another source of uncertainty, which is not of purely EW origin, is the choice
in the parton densities of the factorization scale $M$.
In order to study this dependence, we set $M=\xi\mw$
and consider the canonical range $1/2\leq\xi\leq 2$.
We define the two following relative corrections:
\be
\delta(M) \equiv \frac{\sigma_{\alpha}(M)}{\sigma_{0}(M)} - 1,
~~~~~
\Delta(M) \equiv \frac{\sigma_{\alpha}(M)-\sigma_0(M)}{\sigma_0(\mw)}
\label{deltascale}
\ee
In figure \ref{scaledep} we plot, for the transverse mass
distribution, $\delta(0.5\mw)$ and $\delta(2\mw)$.
The difference between the two curves can be interpreted as mainly
due to the dependence of the \oa cross section on the choice of the QED
factorization scale. We observe a variation at the per mille level of
the transverse mass distribution, as already remarked in ref. \cite{DK}.

In figure \ref{scaledep2} we plot, for the transverse mass
distribution, $\Delta(0.5\mw)$ and $\Delta(2\mw)$.
The difference between the two curves can be interpreted as an
estimate of the uncertainty due to missing
${\cal O}(\alpha\alpha_s)$ corrections, which are of the order of 1\%
around the $W$ resonance.
In fact the numerator in $\Delta(M)$ is proportional to the
hadron-level \oa
corrections, which are a QCD-LO result. A change of the scale in the PDFs
gives an estimate of the QCD-NLO ${\cal O}(\alpha\alpha_s)$ effects
\footnote{
We are not considering in this analysis other contributions of
  the same perturbative order, like the one due to the 
  QED PDFs evolution multiplied
  with the diagrammatic QCD NLO corrections,
  because they are numerically suppressed.}.
We plot in figure 10 the \oa correction at two different scales,
normalized to the Born cross section evaluated at a fixed scale $M=\mw$.
In fact it is known that the introduction of QCD corrections stabilizes the
total cross section against scale variations and we mimic this effect by
setting $M=\mw$ in the denominator of $\Delta(M)$.
In this way the uncertainty due to missing 
${\cal O}(\alpha\alpha_s)$ corrections can be disentangled from other
  purely QCD effects.


To conclude this Section, we would like to comment on the possible relevance of
the different effects and uncertainties on the $W$ boson mass
measurement. It is known in the literature~\cite{CDF,BW,CMNT} 
that the distortion of the
$M_\perp$ distribution around the peak due to the EW \oa and QED
higher-order final-state corrections induces a shift of the extracted
$\mw$ of the order of 100 and 10 MeV respectively (for the muon case).
In the new version of the generator $\tt HORACE$ both effects are
included in a unique tool.

We remark that the theoretical uncertainty due 
to the QCD factorization scale choice, shown in
figure \ref{scaledep2},
 modifies the impact of
the EW radiative corrections by an amount which can be of the order of
1\%: these effects may induce a systematic error in the $\mw$
measurement which can be comparable with the aimed experimental
accuracy and should be carefully considered in future experimental analyses.

\subsection{Rapidity distributions and charge asymmetry}
\begin{figure}
\begin{center}
\includegraphics[height=80mm,angle=0]{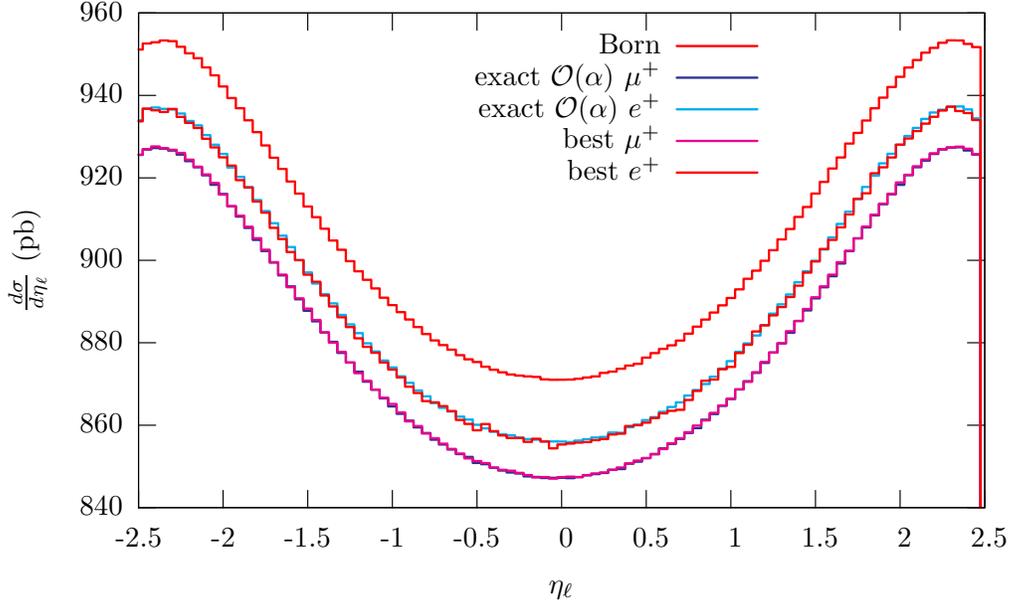}
\end{center}
\caption{Charged lepton pseudo-rapidity distribution in Born, \oa and
  best approximations.}
\label{leptonrapidity1}
\end{figure}

In figure~\ref{leptonrapidity1}
the muon and electron pseudo-rapidity distribution is presented in
the approximations 1., 4. and 5. of table~\ref{tableapprox}.
We observe that the effect of the radiative corrections is almost
constant over the whole range in pseudo-rapidity
and that it is dominated by the \oa term, which gives a correction
negative of approximately -2.7\% in the case of muons and of -1.8\%
for recombined electrons.
Higher-order terms modify the result at the per mille level.

\begin{figure}
\begin{center}
\includegraphics[height=80mm,angle=0]{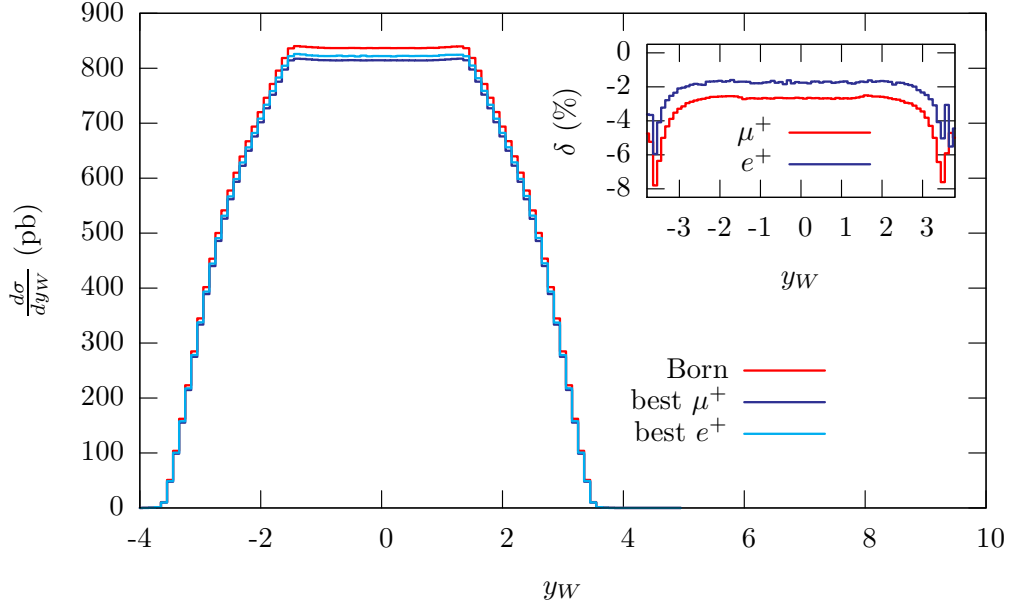} \\
\end{center}
\caption{$W$ boson rapidity distribution in Born, \oa and
  best approximations.}
\label{Wrapidity}
\end{figure}

In figure \ref{Wrapidity} the $W$-boson rapidity is also presented.
With the chosen cuts, this distribution is
essentially flat in the central rapidity interval
$|y_{W}|<1.7$. The radiative corrections are
negative and quite flat and reduce the Born distribution of about -2\%
for the electrons and of -3\% for the muons, as shown in the inset.

\begin{figure}
\begin{center}
\includegraphics[height=45mm,angle=0]{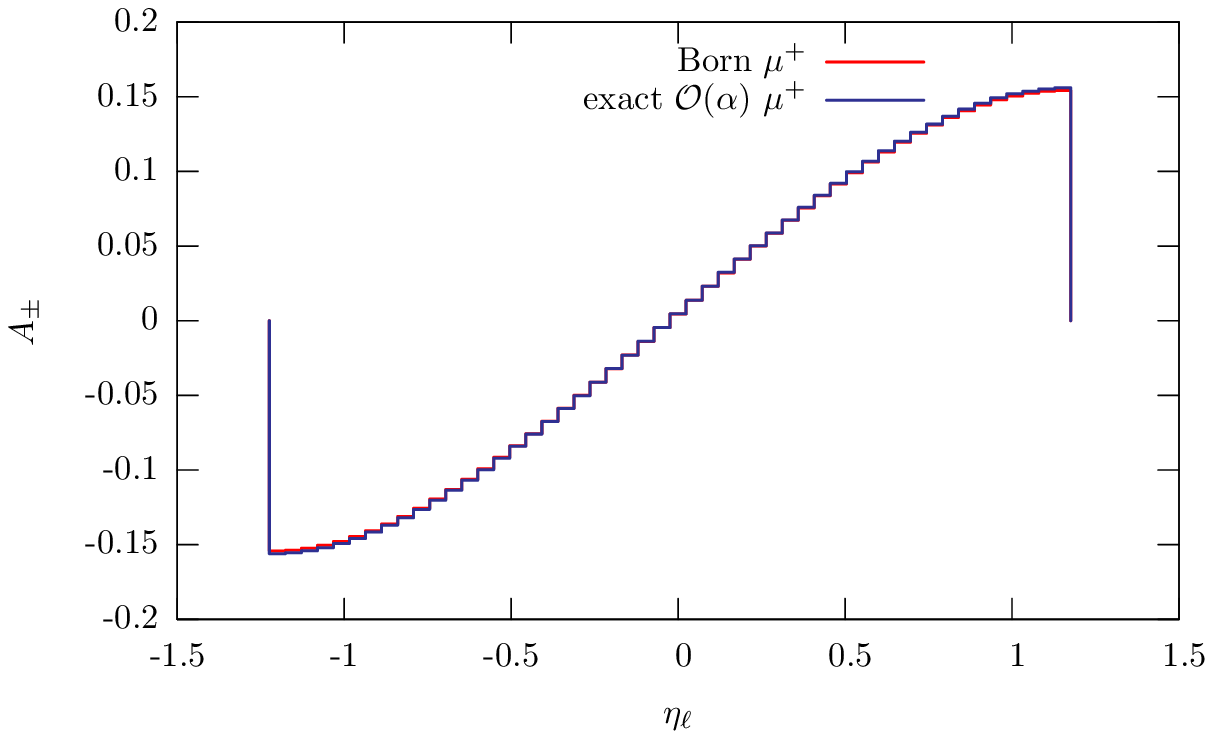}~\includegraphics[height=45mm,angle=0]{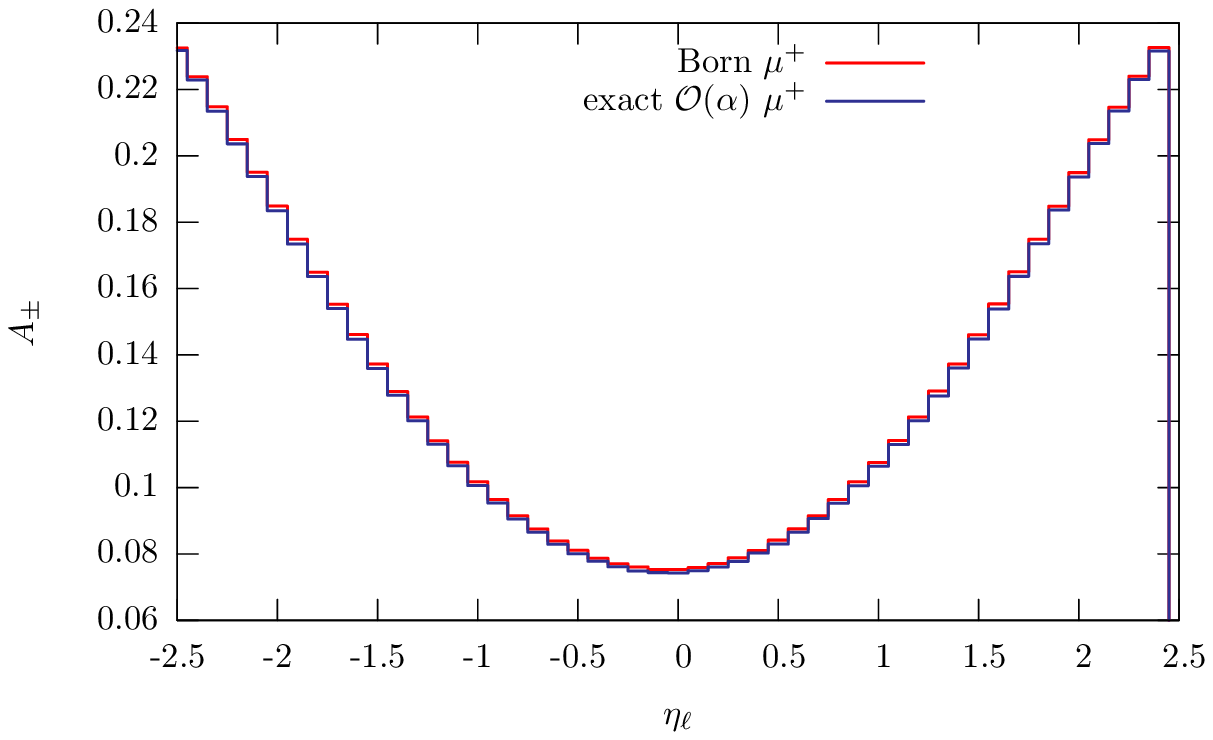}
\end{center}
\caption{Charge asymmetry as function of the muon pseudo-rapidity at
Tevatron ($\sqrt{s}=1.96$ TeV) (left panel) 
and LHC ($\sqrt{s}=14$ TeV) (right panel), in Born and \oa approximation.}
\label{chargeasy}
\end{figure}

The $W$ charge asymmetry presented in figure~\ref{chargeasy}
is defined as
\be
A(\eta_\ell) =
\frac{d\sigma^+/d\eta_\ell~-~d\sigma^-/d\eta_\ell}{d\sigma^+/d\eta_\ell~+~d\sigma^-/d\eta_\ell}
\ee
where $d\sigma^\pm = d\sigma(p\smartpap\to \ell^\pm \nu+X) $;
the asymmetry is due to the partonic content of the incoming hadrons, 
which leads to different lepton pseudo-rapidity distributions in the
production of $W^+$ or $W^-$.
The charge asymmetry can be studied both at the Tevatron and at the LHC,
with different results due to the two different initial states and to
the different ranges of partonic $x$ probed at the two colliders.
At the LHC the function is even under $\eta_\ell\to-\eta_\ell$,
whereas it is odd at the Tevatron. 
The effect of the \oa corrections is at the 1\% level,
while higher-order effects are numerically negligible.


\subsection{$W$ transverse momentum and photonic observables}
\begin{figure}
\begin{center}
\includegraphics[height=80mm,angle=0]{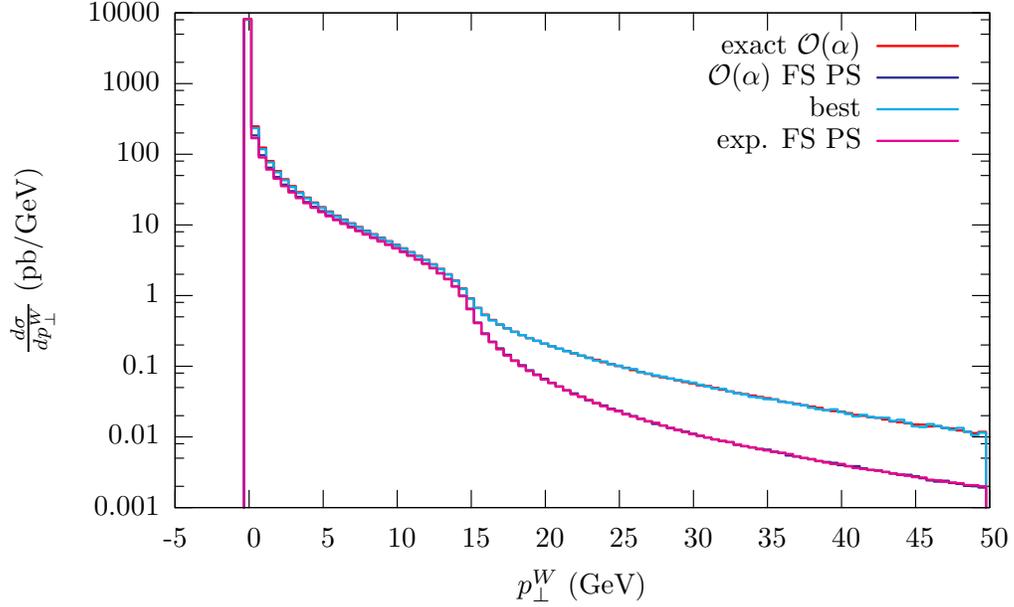}
\end{center}
\caption{
Distribution of the $W$ boson transverse momentum, defined as the
transverse momentum of the final state lepton pair.
}
\label{ptw}
\end{figure}
Real photon radiation gives to the final state lepton pair a
transverse momentum, 
which defines the $W$ boson transverse momentum, whose 
distribution is presented in figure~\ref{ptw}
in the approximations 2., 3., 4. and 5. of table~\ref{tableapprox} 
\footnote{In the present study, the transverse motion of the $W$ boson, as due to
initial-state QCD radiation, is neglected, because it requires a careful inclusion of
QCD corrections, which is beyond the scope of the paper.}.
The large difference in the tail is due 
to a better description of the hard photon radiation
given by the exact \oa matrix element, with respect to its Parton
Shower approximation.
The same comment applies when including, in the two cases,
multiple-photon radiation.

\begin{figure}
\begin{center}
\includegraphics[height=80mm,angle=0]{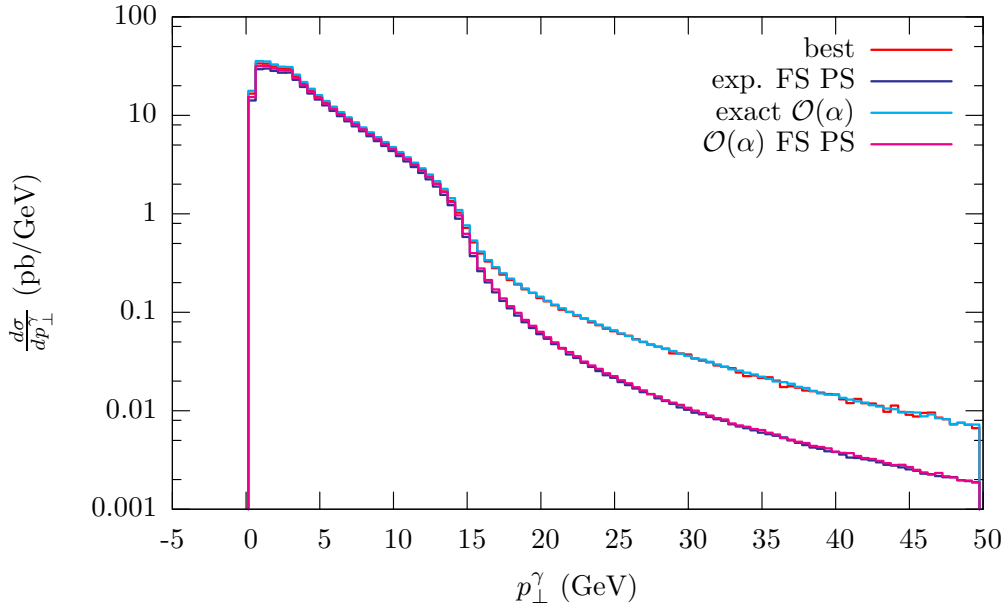}
\end{center}
\caption{
Distribution of the hardest photon transverse momentum.
}
\label{ptg}
\end{figure}

\begin{figure}
\begin{center}
\includegraphics[height=80mm,angle=0]{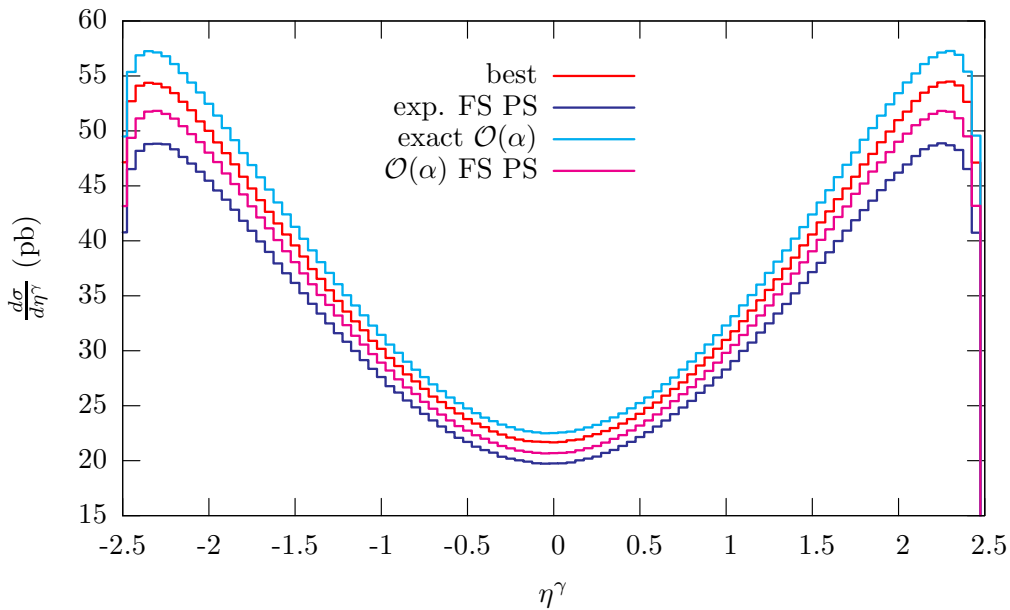}
\end{center}
\caption{
Distribution of the hardest photon pseudorapidity.
}
\label{etag}
\end{figure}

We now present some distributions for the
radiative Drell-Yan event, with a hard photon associated to the
large transverse momentum lepton pair. 
This signature can be a source of information
to study for instance the trilinear gauge boson
$WW\gamma$ vertex \cite{CDFtrilinear}.
We select the events by imposing the cuts of eq.~\myref{cuts} and
requiring that the most energetic photon is detected, i.e.
with rapidity $|\eta_\gamma|<2.5$ and $E_\gamma>3$ GeV.
The transverse momentum and the rapidity distributions
of the hardest photon
are plotted in figures~\ref{ptg} and~\ref{etag}. 
As in the $W$ transverse momentum case,
the large difference in the tail in figure~\ref{ptg} is due 
to the better description of the hard photon radiation
given by the exact \oa matrix element.
Concerning the hardest photon rapidity distribution in figure
\ref{etag},
we compare first the lowest order results, which have been obtained in
two different approximations, namely 2. and 3. of table
\ref{tableapprox}
and observe the increase of the cross section given by the exact
matrix element.
When including multiple-photon radiation, the lowest order results
receive a negative correction.

\section{Conclusions}
\label{concllabel}
In this paper we presented a precision calculation of 
the charged current Drell-Yan process,
which includes both the exact EW \oa and the
leading-log multiple-photon corrections.
In order to keep consistently 
under control the two effects, a matching algorithm between the fixed
order calculation and a QED Parton Shower, avoiding the double
counting of the leading-log corrections, has been devised
and implemented. To our knowledge, the matching algorithm here
presented is the first example of such an application in the field
of EW radiative corrections. Initial-state QED collinear singularities have been
regularized by means of finite quark masses, requiring a subtraction
of the initial-state logarithms (in analogy to QCD NLO
calculations) which are already accounted for in the evolution of the
PDFs. 
The subtraction procedure, already known at \oab,
has been generalized to the QED resummed cross section, 
to systematically remove initial-state
logarithms at all orders. Our results have been verified to be
completely independent of the value of the quark masses, as expected.

We studied,
with the new version of the event generator $\tt HORACE$, 
the impact of different classes of radiative corrections on
several physical observables, showing
the importance of combining in a unique tool
fixed order results with the resummation of multiple-photon radiation.
In fact, radiative corrections induce
effects ranging from several per mille to few per cent of the lowest-order
cross section; they change the integrated cross section and the shape
of the distributions.

For example, 
the value of the $W$ boson mass,
extracted from the transverse mass distribution,
is shifted, in the case of a final-state muon, 
by EW \oa corrections of $ {\cal O}$(100~MeV)
and of $ {\cal O}$(10~MeV) by multiple photon emission.
These effects are important in view of the
measurement foreseen at the LHC and Tevatron with an accuracy of
$\Delta\mw \approx 15$  and $\approx 30$ MeV and
can now be included in a systematic way in the analysis of the
experimental data.
Far from the Jacobian peak, the tail of the transverse mass
distribution, where the Drell-Yan process is a background to
new gauge boson searches, receives large negative
corrections due to the presence of the \oa EW Sudakov logarithms, of the
order of 15-30\%.
The effect of EW corrections is important also for the estimate of the
detector acceptances in view of exploiting the $W$ production process
as a precise luminosity monitor at the LHC, with an accuracy of some
per cent.

We also studied radiative events, where at least one hard photon
accompanies the final
state lepton pair, which are a useful source of information for instance
to measure the trilinear gauge boson vertex.
In view of precision studies,
the effects of the
radiative corrections and of treating with the exact matrix element
the photon radiation
are sizeable on observables like the photon transverse momentum and rapidity
distributions.

Besides the EW \oa and multiple-photon corrections, we studied some of
the remaining theoretical uncertainties. We considered the ones due to 
the EW input scheme choice and to the QCD factorization scale choice.
We would like to remark
that they should be considered in detail to quantify, for example, the
systematic error on the $W$ mass measurement.
%

A possible development of the work presented in this paper is the 
implementation in $\tt HORACE$ of the complete EW \oa corrections to
the $Z$ production process and a detailed phenomenological
study of their impact on
the neutral current Drell-Yan process, which is important for many
measurements and calibrations at hadron colliders. Furthermore, we are
now working to
combine the results presented here with QCD corrections, aiming at
providing a unified tool which includes the relevant EW and QCD
effects and which can be very useful for the experimental collaborations.

\vskip 24pt
\leftline{\bf Acknowledgments}
\vskip 12pt\noindent
We are grateful to Andreij Arbuzov, Dima Bardin, Ulrich Baur, Stefan
Dittmaier, Michael Kr\"amer
and Doreen Wackeroth for their precious collaboration during the tuned
comparisons 
of the 2005 Les Houches workshop ``Physics at TeV colliders''. We are
indebted with Mauro Moretti
for his help in the comparison with the results of the $\tt ALPHA$ code and
with Massimiliano Bellomo,
Fulvio Piccinini and Giacomo Polesello for useful discussions and
interest in our work.
We thank Stefano Forte for a careful reading of the preliminary manuscript.


%

\end{document}